\documentclass[twocolumn]{aastex62}

\shorttitle{Orbital Refinement and Stellar Properties}
\shortauthors{Michelle Hill et al.}
\submitjournal{\aj}
\accepted{Mar 04, 2020}
\usepackage{footnote}
\usepackage{float}
\usepackage{graphicx,textcomp,fancyhdr,hyperref}
\usepackage[noabbrev]{cleveref}
\usepackage{footnote}
\usepackage{url}
\usepackage[flushleft]{threeparttable}
\usepackage[caption=false]{subfig}
\begin{document}

\title{Orbital Refinement and Stellar Properties for the HD~9446, HD~43691, and HD~179079 Planetary Systems}

\author[0000-0002-0139-4756]{Michelle L. Hill}
\affiliation{Department of Earth and Planetary Sciences, University of California, Riverside, CA 92521, USA}
\email{michelle.hill@email.ucr.edu}

\author{Teo Mo\v{c}nik}
\affiliation{Department of Earth and Planetary Sciences, University of California, Riverside, CA 92521, USA}

\author{Stephen R. Kane}
\affiliation{Department of Earth and Planetary Sciences, University of California, Riverside, CA 92521, USA}

\author{Gregory W. Henry}
\affiliation{Center of Excellence in Information Systems, Tennessee State University, Nashville, TN 37209, USA}

\author{Joshua Pepper}
\affiliation{Department of Physics, Lehigh University, Bethlehem, PA 18015, USA}

\author{Natalie R. Hinkel}
\affiliation{Southwest Research Institute, San Antonio, TX 78238, USA}

\author[0000-0002-4297-5506]{Paul A. Dalba}
\altaffiliation{NSF Astronomy and Astrophysics Postdoctoral Fellow}
\affiliation{Department of Earth and Planetary Sciences, University of California, Riverside, CA 92521, USA}

\author{Benjamin J. Fulton}
\affiliation{Department of Astronomy, California Institute of Technology, Pasadena, CA 91125, USA}

\author{Keivan G. Stassun}
\affiliation{Vanderbilt University, Department of Physics \& Astronomy, 6301 Stevenson Center Lane, Nashville, TN 37235, USA}

\author{Lee J. Rosenthal}
\affiliation{Department of Astronomy, California Institute of Technology, Pasadena, CA 91125, USA}

\author{Andrew W. Howard}
\affiliation{Department of Astronomy, California Institute of Technology, Pasadena, CA 91125, USA}

\author[0000-0002-2532-2853]{Steve B. Howell}
\affiliation{NASA Ames Research Center, Moffett Field, CA 94035, USA}

\author{Mark E. Everett}
\affiliation{National Optical Astronomy Observatory, Tucson, AZ 85719, USA}

\author[0000-0001-9879-9313]{Tabetha S. Boyajian}
\affiliation{Department of Physics and Astronomy, Louisiana State University, Baton Rouge, LA 70803, USA}

\author{Debra A. Fischer}
\affiliation{Department of Astronomy, Yale University, New Haven, CT 06511, USA}

\author[0000-0001-8812-0565]{Joseph E. Rodriguez}
\newcommand{\cfa}{Center for Astrophysics \textbar \ Harvard \& Smithsonian, 60 Garden St, Cambridge, MA 02138, USA}

\author[0000-0002-9539-4203]{Thomas G. Beatty}
\affiliation{Department of Astronomy and Steward Observatory, University of Arizona, Tucson, AZ 85721}

\author[0000-0001-5160-4486]{David J. James}
\affiliation{Center for Astrophysics $\vert$ Harvard \& Smithsonian, 60 Garden Street, Cambridge, MA 02138, USA}
\affiliation{Black Hole Initiative at Harvard University, 20 Garden Street, Cambridge, MA 02138, USA}

\begin{abstract}

The Transit Ephemeris Refinement and Monitoring Survey (TERMS) is a project which aims to detect transits of intermediate-long period planets by refining orbital parameters of the known radial velocity planets using additional data from ground based telescopes, calculating a revised transit ephemeris for the planet, then monitoring the planet host star during the predicted transit window. Here we present the results from three systems that had high probabilities of transiting planets: HD~9446~b~\&~c, HD~43691~b, \& HD~179079~b. We provide new radial velocity (RV) measurements that are then used to improve the orbital solution for the known planets. We search the RV data for indications of additional planets in orbit and find that HD~9446 shows a strong linear trend of 4.8$\sigma$. 
Using the newly refined planet orbital solutions, which include a new best-fit solution for the orbital period of HD~9446~c, and an improved transit ephemerides, we found no evidence of transiting planets in the photometry for each system.  Transits of HD~9446~b can be ruled out completely and transits HD~9446~c \& HD~43691~b can be ruled out for impact parameters up to b = 0.5778 and  b = 0.898 respectively due to gaps in the photometry. A transit of HD~179079~b cannot be ruled out however due to the relatively small size of this planet compared to the large star and thus low signal to noise. We determine properties of the three host stars through spectroscopic analysis and find through photometric analysis that HD~9446 exhibits periodic variability.

\end{abstract}


\bigskip

\section{Introduction}
\label{introduction}

Transiting exoplanets have become an integral part of exoplanetary science and provide not only detection and confirmation of exoplanets but also information about the radius and the atmospheric composition of the planet. Provided the star's size is known, the radius of the planet can be determined by the reduction in a star's flux as the planet passes in front of the star since the transit ``depth" (or corrected depth if a close companion provides ``third light” contamination) is directly proportional to the cross-sectional size of the planet.
The atmospheric composition of the planet can be detected by the light from the star passing through the planet's atmosphere, through a method called transmission spectroscopy \citep{seager2000}. 

Combined with the information gathered through the radial velocity (RV) method, transits enable studies of the mean density of a planet as well as its interior structure. Planets with both RV and transit observations are the most fully constrained and are the main data sources behind many statistical relationships that define our understanding of the formation and evolution of exoplanets, such as the mass-radius relationship and models of the composition of exoplanets. Thus detecting the potential transits of known RV planets is of great importance. 

The Transit Ephemeris Refinement and Monitoring Survey (TERMS) is a project that seeks to detect transits of known RV planets with a broad range of orbital periods \citep{Kane09}. The transit detection method is biased towards planets with short orbital periods close to their star as the probability of a planet transiting its star increases with decreasing distance from the star \citep{Beatty08,Kipping16}. In order to detect planets further away from their star and broaden the catalog of transiting exoplanets, the TERMS team targets known intermediate to long period exoplanets with existing RV data in the literature. The orbital parameters of the known planets are first re-calculated using the latest data from Keck Observatory. Then a refined transit ephemeris is determined for the planet so that the planet host star can be observed during the predicted transit window. Photometry from various ground based telescopes including two of Tennessee State University's automated photoelectric telescopes (APTs) and the Kilodegree Extremely Little Telescope (KELT) are then analysed to determine if a transit was detected. 

In this paper we present an analysis of both the host star and the planets in the HD~9446, HD~43691 \& HD~179079 systems. We provide new RV data for each system, extending the time baseline of the observations and so improving the
orbital parameters of the planets. From this new data, we calculated
an accurate transit ephemeris and predicted the parameters of
a potential transit of each known planet. We present precision photometry of HD~9446, HD~43691 \& HD~179079 acquired with the APTs and KELT and determine the existence of transits of each of the known planets in these systems. We note that each of these stars falls within the gaps of the primary mission for TESS and so are unlikely to be observed by space based instruments in the near future.  


\section{Science Motivation}
\label{SM}

\begin{figure*}
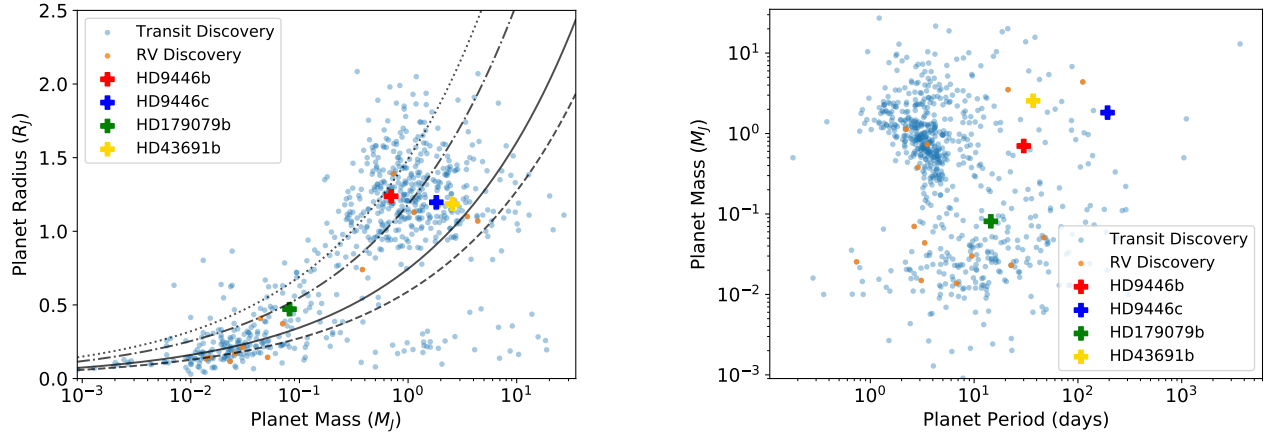

\gridline{\fig{"Mass_Radius_new"}{0.45\textwidth}{}
          \fig{"Mass_period_new"}{0.45\textwidth}{}}
\caption{The location of the studied planets with respect to the known demographics of exoplanets. Left: Mass-radius diagram of known transiting planets (small circles), including lines of constant density. The four planets included in this study are shown as large crosses, where the radii are estimated using the \citet{Chen16} methodology (see Section~\ref{TP}). Right: Period-Mass diagram using the same planets as for the mass-radius diagram where the planets in our sample are similarly indicated as crosses.}
\label{fig:Mass_Rad_per}
\end{figure*}

Each of the systems in this paper was detected by the RV method. The multi-planet system HD~9446 was first discovered by \citet{Hebrard10} using the SOPHIE spectrograph. The initial orbital solution gave the inner planet (b) a period of $30.052\pm0.027$ days and the outer planet (c) a period of $192.9\pm0.9$~days. HD~43691~b  was first discovered by \citet{Dasilva07}, with an orbital parameters updated in 2018 by \citet{Ment18} including a refined orbital period for planet b of $36.9987\pm0.0011$~days. HD~179079~b was first discovered by \citet{Valenti09}, with updated orbital parameters including a period of $14.479\pm0.010$~days by \citet{Ment18} (see Table \ref{tab:Stellar}).	
As the precision of planetary and orbital parameters is inherently linked to how well the host star is known we present an SED analysis of each star in Section \ref{SP}.
Each of these systems have a planet that occupies the period space between 10 and 50 days and have relatively high transit probabilities (see Section~\ref{TP}) so were flagged as high-priority TERMS targets. Shown in Figure~\ref{fig:Mass_Rad_per} are two diagrams that include the known transiting planets as small circles, where the data were extracted from the NASA Exoplanet Archive \citep{Akeson2013} on December 1, 2019. Highlighted as orange circles are those transiting planets discovered using the RV method, such as HD~209458b \citep{Charbonneau2000,Henry2000}. The four planets in the three systems included in this study are indicated as large crosses. Note that these four planets are not known to transit, and so we use the radii estimates described in Section~\ref{TP}. The mass-radius diagram shown in the left panel indicates where the four planets fall with respect to the bulk of the known transiting planet population. HD~179079~b is expected to lie within the Neptune regime and could potentially be an interesting test of photo-evaporation for low-mass giant planets \citep{Laughlin2011,Lopez2013}. The remaining three planets fall within the bulk of the known transiting giant planets. However, the period-mass diagram shown in the right panel demonstrates that the three largest planets in our sample are significantly removed from the bulk of the known transiting giant planet population when considering orbital period. The detection of transits for such planets would therefore be of high value in studies of mass-radius relationships for "cold" giant planets, particular since the host stars for the three systems of our sample are all exceptionally bright (see Section~\ref{SP}).

In order to detect the possible transit of the planetary systems, additional RV measurements were taken using the HIRES instrument on the Keck I telescope. Combining the newly obtained data with that already available for each system, the RV data was fit using the RV fitting tool RadVel \citep{Fulton18} to confirm each planet's orbital solution and look for linear trends (Section \ref{RVA}). These trends could potentially be indications of additional unidentified planets in the system. We search for unknown planets in the RV data and provide an analysis of the sensitivity of our search through RV injection-recovery tests in Section \ref{IT}. 

Using the transit window determined through RadVel along with the photometry from the KELT and the APTs, the newly calculated transit windows were then used to look for transits of each planet. These results are presented in Section \ref{SfT}.

The KELT and APT photometry were then used to characterize the stellar variability of each system. Starspots are a common cause of intrinsic stellar variability by inducing rotational modulation as the star rotates (e.g. \citet{Kipping12b} and citations therein).
The rotational modulation period is therefore a direct measurement of the rotational period of the star. When coupled with the spectroscopic measurement of the projected rotational velocity, it can be used to constrain the inclination of the stellar rotational axis, $i_*$.

\section{Stellar Properties}
\label{SP}

\begin{figure}
\centering 
\subfloat{%
  \includegraphics[width=0.9\columnwidth]{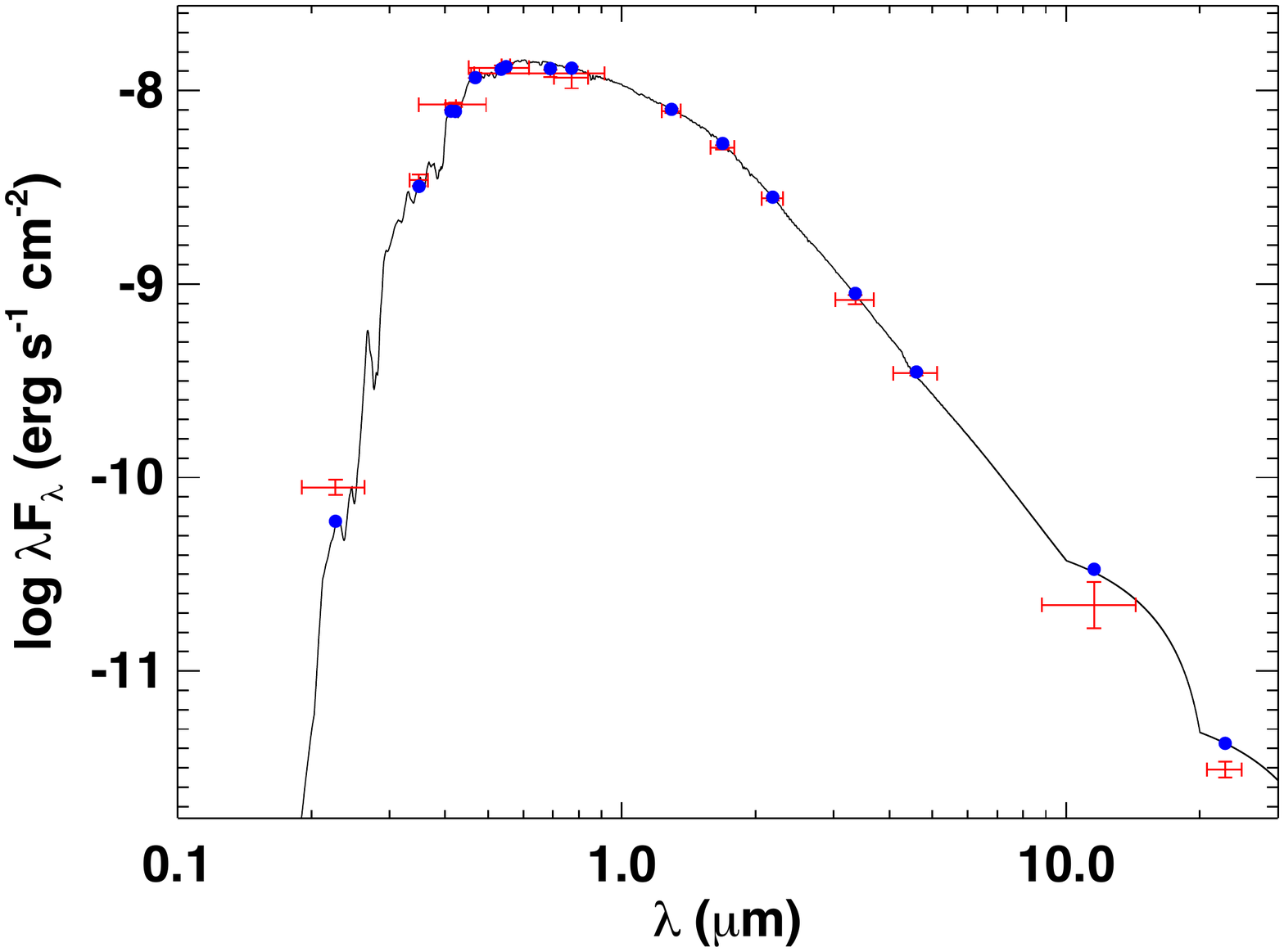}%
}\qquad
\subfloat{%
  \includegraphics[width=0.9\columnwidth]{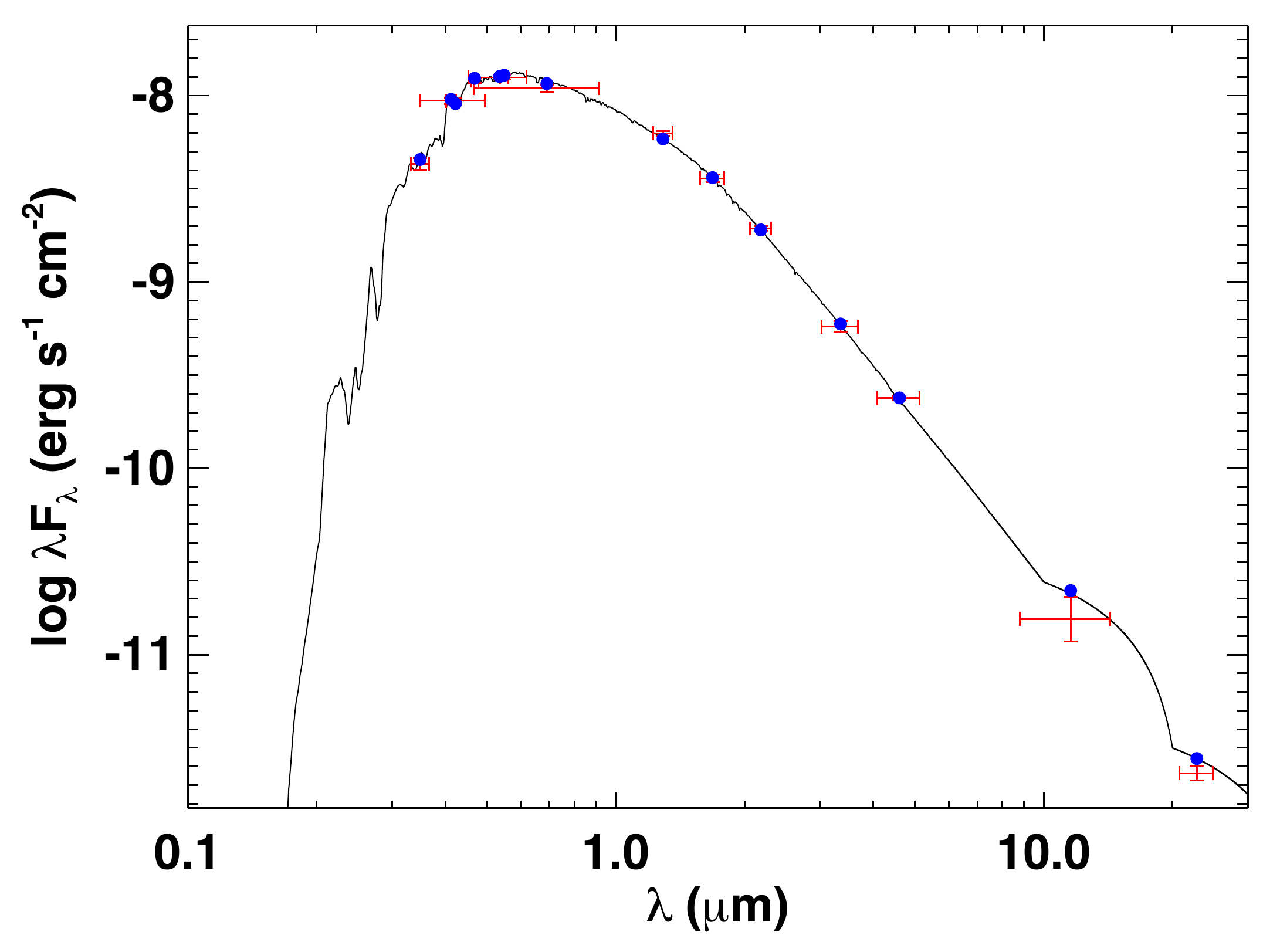}%
}\qquad
\subfloat{%
  \includegraphics[width=0.9\columnwidth]{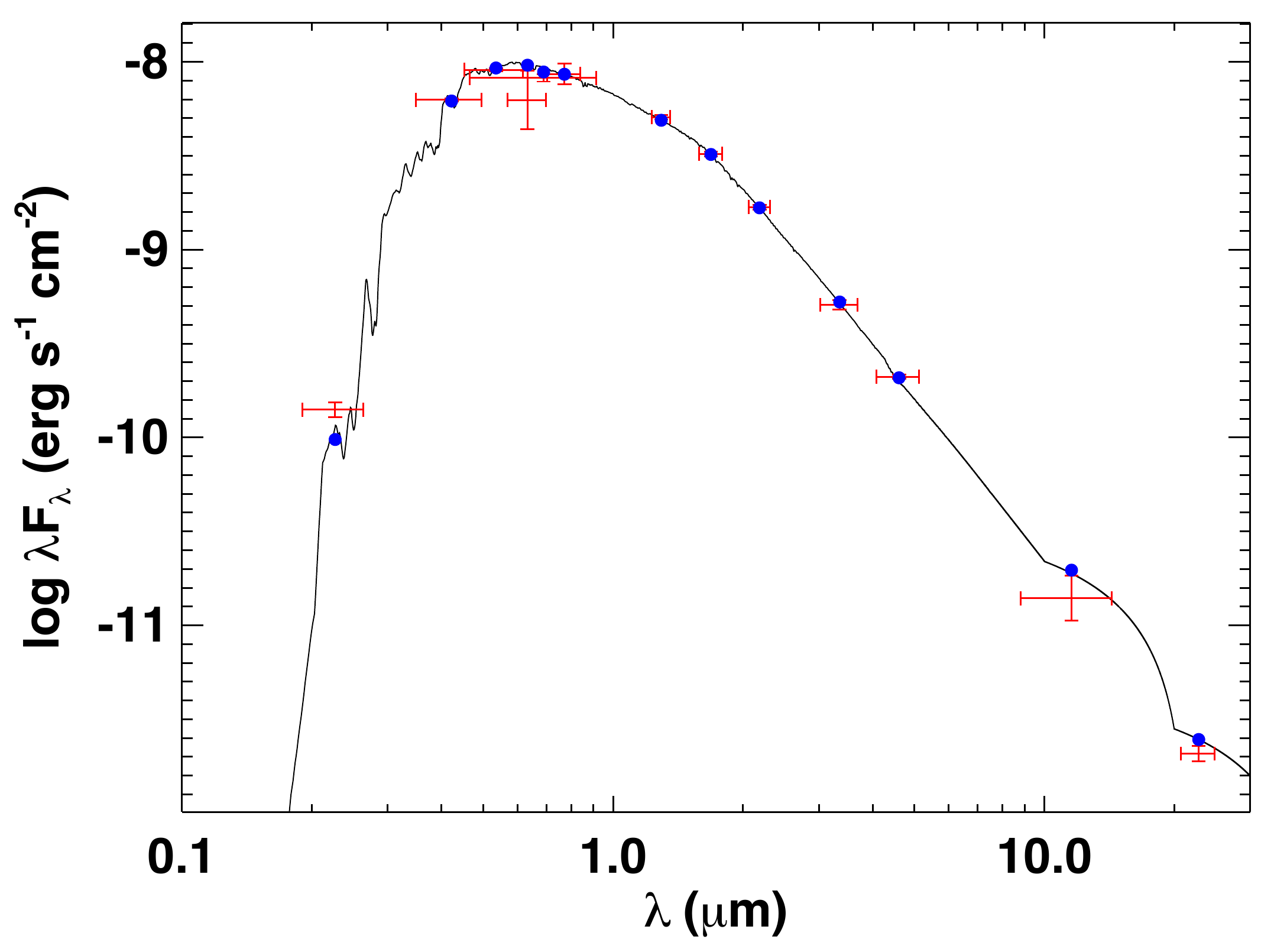}%
}
\caption{Spectral energy distributions (SEDs) of (a) HD~179079 with $\chi^{2}$ value of 3.4, (b) HD~43691 $\chi^{2}$ value of 1.9, and (c) HD~9446 $\chi^{2}$ value of 2.0. Red symbols represent the observed photometric measurements, where the horizontal bars represent the effective width of the passband. Blue symbols are the model fluxes from the best-fit Kurucz atmosphere model (black).  \label{fig:sed}}
\end{figure}

As an independent determination of the basic stellar parameters, we performed an analysis of the broadband spectral energy distribution (SED) of each star together with the {\it Gaia\/} DR2 parallaxes \citep[adjusted by $+0.08$~mas to account for the systematic offset reported by][]{StassunTorres2018}, in order to determine an empirical measurement of the stellar radius, following the procedures described in \citet{Stassun2016,Stassun2017,Stassun2018}. We extracted the NUV flux from {\it GALEX}, the $B_T V_T$ magnitudes from {\it Tycho-2}, the Str\"{o}mgren $ubvy$ magnitudes from \citet{Paunzen:2015}, the $BVgri$ magnitudes from APASS, the $JHK_S$ magnitudes from {\it 2MASS}, the W1--W4 magnitudes from {\it WISE}, and the $G$ magnitude from {\it Gaia}, as available for each star. Together, the available photometry in general spans the full stellar SED over the wavelength range 0.2--22~$\mu$m (see Figure~\ref{fig:sed}). 

We performed a fit using Kurucz stellar atmosphere models \citep{Kurucz13}, with the effective temperature ($T_{\rm eff}$) and metallicity ([Fe/H]) and surface gravity ($\log g$) adopted from the systematic spectroscopic analyses provided in the SWEET catalog \citep{Santos:2013}. The only free parameter is the extinction ($A_V$), which we restricted to the maximum line-of-sight value from the dust maps of \citet{Schlegel:1998}. The resulting fits are good (Figure~\ref{fig:sed}) with a reduced $\chi^2$ ranging from 1.9 to 3.4. Integrating the (unreddened) model SED gives the bolometric flux at Earth ($F_{\rm bol}$). Taking the $F_{\rm bol}$ and $T_{\rm eff}$ together with the {\it Gaia\/} DR2 parallax, gives the stellar radius. Finally, we can use the empirical relations of \citet{Torres:2010} and a 6\% error from the empirical relation itself to estimate the stellar mass; this, in turn, together with the stellar radius provides an empirical estimate of the mean stellar density. We used the derived radius and mass to calculate a value for $\log g$ and compared our values to those from the SWEET catalog \citep{Santos:2013}. We found there is a good agreement (within the uncertainties) between the values for each star.
The full set of resulting parameters are summarized in Table~\ref{tab:Stellar} along with planet properties of the three systems drawn from the \citet{Hebrard10} and \citet{Ment18} papers. Note that while HD~9446 and HD~43691 are similar to the Sun, HD~179079 is an evolved star that has entered the sub-giant branch.

We observed the star HD~9446 using CHIRON, a fiber-fed Echelle spectrometer with a resolution of $R = 79,000$ \citep{tokovinin2013,brewer2014} in use at the 1.5-m telescope at Cerro Tololo Inter-American Observatory (CTIO). The CHIRON spectrum was analyzed using Spectroscopy Made Easy (SME), described in detail by \citet{valenti1996} and \citet{valenti2005}. The stellar parameters derived for HD~9446 from our SME analysis are consistent with those shown in Table~\ref{tab:Stellar}. The distances to the stars shown in Table~\ref{tab:Stellar} are from the Gaia Data Release 2 parallaxes \citep{Gaia18}. 

The three host stars were found within the Hypatia Catalog\footnote{Data can be found on \url{www.hypatiacatalog.com}.}, an amalgamate database of stellar abundances for nearby FGKM-type stars \citep{Hinkel14}. HD~9446 was observed by 7 groups \citep{Ramirez07, Ramirez09, Baumann10, Brugamyer11, Ramirez12a, Ramirez13, Brewer16}, while HD~43691 and HD~179079 were both observed by 6 groups (\citealt{Gonzalez10a,Gonzalez10b,Brugamyer11,Kang11,Petigura11,Brewer16} and \citealt{Petigura11,Maldonado13,Ramirez14a,Jofre15,Brewer16,Maldonado16}, respectively). All datasets are renormalized to the same solar scale within Hypatia, namely \citet{Lodders09}, such that they are more easily comparable. Then, when multiple groups measure the same element within the same star, the median value is used and the uncertainty is defined as the {\it spread} or range from all of the renormalized abundance determinations. The [Fe/H] values for the three target stars are consistent with those in Table~\ref{tab:Stellar}. The other elements, including volatiles, refractory, iron-peak, and neutron-capture, are all enriched to a similar magnitude as the respective [Fe/H] values when compared to the Sun. Of note, the composition of HD~9446 is the most similar to the Sun, such that a few elements (N, O, Na, Mg, Si, S, Mn, Cu) hover around 0.0 dex when uncertainty is taken into account. All three stars are relatively close and bright, making the systems ideal for additional characterization.

We also examine the stars for possible evidence of binary companions. HD~9446 was previously observed using speckle imaging by \citet{wittrock2016}, for which the data found no evidence of a stellar companion. Here we provide new results for one of the stars, HD~179079, using data from the Differential Speckle Survey Instrument (DSSI). DSSI is a dual-channel speckle imaging system that uses two narrowband filters with wavelengths centered on 692 and 880~nm. The details of the instrument hardware and data reduction are described in detail by \citet{horch2009} and \citet{Howell11}. HD~179079 was observed using DSSI as part of a survey of known exoplanet host stars for binary companions \citep{kane2014,wittrock2016, wittrock2017,kane2019}. The star was observed using WIYN on April 19, 2016 and Gemini-South on June 29, 2016, during which periods DSSI was mounted as a guest instrument at these telescopes. No companion was detected in any of the images from our observations, with a detection limit of $\Delta m \sim 7$ magnitudes in both channels within the range of 0.1--1.8$\arcsec$. Given the distance to the star listed in Table~\ref{tab:Stellar}, this corresponds to a projected range of 7--125~AU for which companions of most spectral types can be excluded based on our analysis. 

\begin{deluxetable*}{lllll}
\tablecaption{Stellar Properties \label{tab:Stellar}}
\tablehead{
  \colhead{Parameter} & 
  \colhead{HD 9446 b\tablenotemark{a}} & 
  \colhead{HD 9446 c\tablenotemark{a}} & 
  \colhead{HD 43691 b\tablenotemark{b}} &
  \colhead{HD 179079 b\tablenotemark{b}}
}
\startdata
\sidehead{\bf{Star}}
Spectral Type & G5 V & G5 V & G0 & G5IV \\
$V$ & 8.35 & 8.35 & 8.03 & 7.95 \\
Distance (pc)\tablenotemark{c}
& $50.42\pm0.35$ & $50.42 \pm 0.35$ & $85.81\pm0.37$ & $69.848\pm0.395$ \\
$T_\mathrm{eff}$ (K)\tablenotemark{d} & $5793\pm22$ & $5793\pm22$ & 6093 & 5646 \\
$\log g$\tablenotemark{d} & $4.53\pm0.16$ & $4.53\pm0.16$ & $4.31\pm0.07$ & $4.11\pm0.04$ \\
$F_{\rm bol}$ ($10^{-8}erg s^{-1}cm^{-2}$)\tablenotemark{d} & $1.245\pm0.029$ & $1.245\pm0.029$ & $1.764\pm0.020$ & $2.084\pm0.049$ \\
$[$Fe/H$]$ (dex)\tablenotemark{e} & $0.13 \pm 0.06$ & $0.13\pm0.06$ & $0.32 \pm 0.03$ & $0.35 \pm 0.09$ \\
$M_* \,(M_\odot)$\tablenotemark{f} & $1.07\pm0.08$ & $1.07\pm0.08$ & $1.32\pm0.09$ & $1.25\pm0.09$ \\ 
$R_* \,(R_\odot)$\tablenotemark{f} & $0.984\pm0.01$ & $0.984\pm0.01$ & $1.704\pm0.023$ & $1.792\pm0.016$ \\
$\rho_* (g/cm^3)$\tablenotemark{f} & $1.59\pm0.12$ & $1.59\pm0.12$ & $0.375\pm0.031$ & $0.307\pm0.023$ \\
\sidehead{\bf{Planet}}
$P$ (days) & $30.052\pm0.027$ & $192.9\pm0.9$ & $36.9987\pm0.0011$ & $14.479\pm0.010$ \\
$e$ & $0.2\pm0.06$ & $0.06\pm0.06$ & $0.0796\pm0.0067$ & $0.0490\pm0.0870$ \\
$\omega~(\degr)$ & $215\pm30$ & $100\pm130$ & $292.7\pm4.8$ & $ 308\pm77$ \\
$K$~(m s$^{-1}$) & $46.6 \pm 3.0$ & $63.9 \pm 4.3$  & $130.06\pm0.84$ & $6.22\pm 0.78$ \\
$M_p \sin i~(M_J)$ & $0.7\pm0.06$ & $1.82\pm0.17$ & $2.55\pm0.34$ & $0.081\pm0.02$ \\
$a$ (au) & $0.189\pm0.006$ & $0.654\pm0.022$ & $0.238\pm0.015$ & $0.1214\pm0.0068$ \\
\enddata
\tablenotetext{a}{Planet parameters from \citet{Hebrard10}}
\tablenotetext{b}{Planet parameters from \citet{Ment18}}
\tablenotetext{c}{\citet{Gaia18}}
\tablenotetext{d}{SWEETcat catalog \citet{Santos:2013}}
\tablenotetext{e}{Hypatia Catalog \citet{Hinkel14}}
\tablenotetext{f}{Our SED analysis}
\end{deluxetable*}


\section{Transit Probability}
\label{TP}

As TERMS targets, these planets had a relatively high likelihood of transiting their host star. Using the equation from \citet{Kane08} the prior transit probability can be estimated by:
\begin{equation}
P_t = \frac{(R_p + R_*)[1+e\cos(\pi/2 - \omega)]}{a(1-e^2)}
\label{eq:transprob}
\end{equation}
Where $R_*$ is the radius of the star, $e$ is the orbital eccentricity, $\omega$ is the argument of periastron, $a$ is the semi-major axis and the planet radius ($R_p$) is determined using the mass-radius relation code Forecaster developed by \citet{Chen16}. Our calculation does not compute the posterior transit probability, which includes prior information for the mass distribution of RV-discovered exoplanets \citep{Ho2011, Stevens2013}. Simulations by \citet{Stevens2013} found only a $\sim$1\% difference between prior and posterior transit probabilities for planets with masses in the range 100--1000~$M_{\earth}$, which likely includes HD~9446~b, HD~9446~c, and HD~43691~b. For HD~179079~b, the posterior transit probability is likely higher than the prior transit probability by $\sim$20\% \citep{Stevens2013}. 

Calculating the transit probability (Equation~\ref{eq:transprob}) with the planet radii estimated by Forecaster \citep{Chen16} we find that HD~9446~b, with an estimated planet radius of $1.238\pm0.240$~$R_J$, has a transit probability of $\sim2.2\%$ with a transit depth of $1.671\times10^{-2}\pm7.1\times10^{-3}$. For HD~9446~c, with a radius of $1.197\pm0.226$~$R_J$, the transit probability is $\sim0.7\%$ with an expected transit depth of $1.563\times10^{-2}\pm6.4\times10^{-3}$. For HD~179079~b, with a calculated radius of $0.471\pm0.1935$~$R_J$, the transit probability is $\sim6.6\%$ with a depth of $7.29\times10^{-4}\pm7.2\times10^{-4}$, and for HD~43691~b, with a calculated planet radius of $1.185\pm0.218$~$R_J$, transit probability is $\sim3.1\%$ with a depth of $5.107\times10^{-3}\pm2.05\times10^{-3}$.

\begin{figure}
\centering 
  \includegraphics[width=1.0\columnwidth]{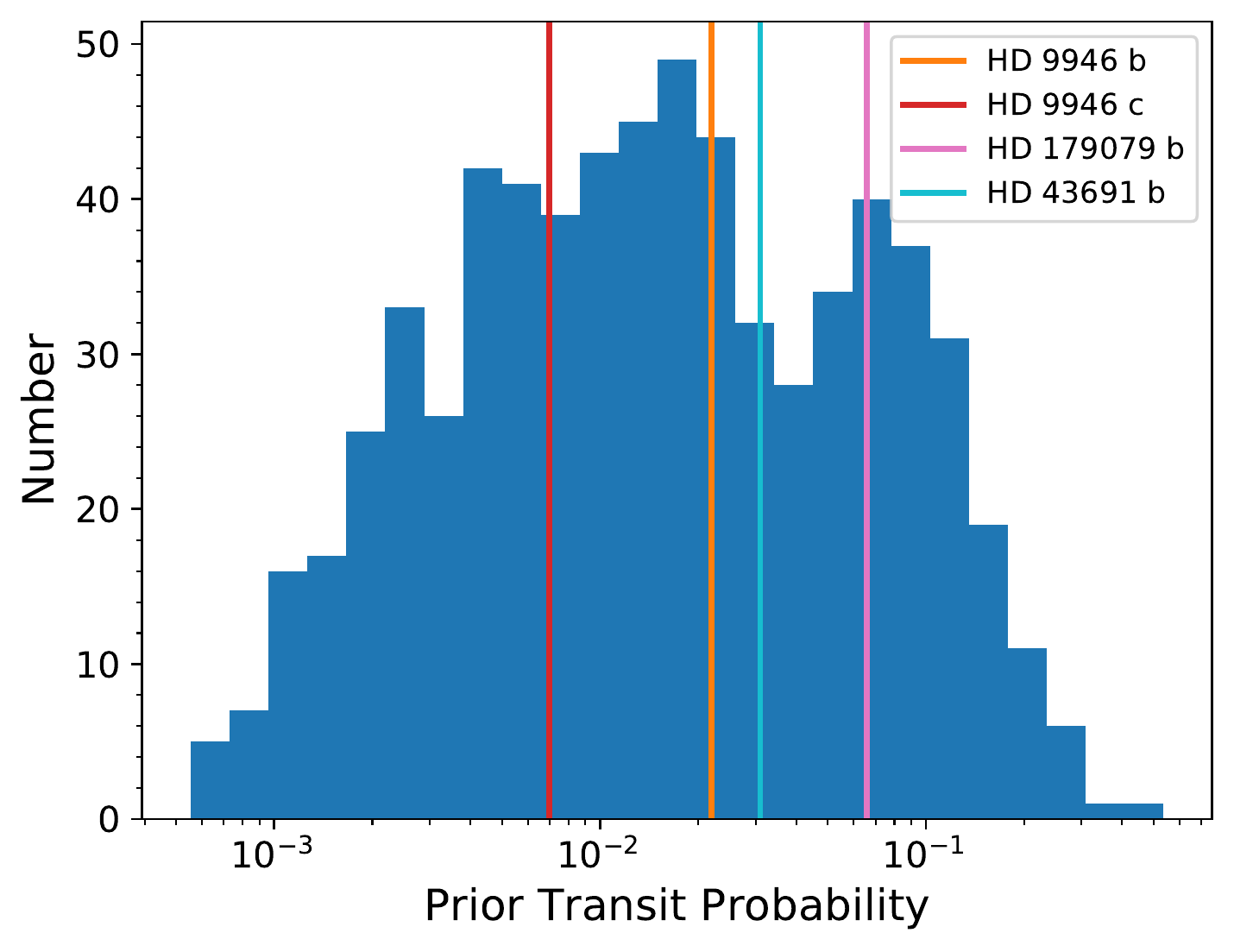}
\caption{The prior transit probabilities for the four target planets compared to the distribution of prior transit probabilities for RV-discovered planets from \citet{Dalba19}.}
 \label{fig:trans_prob}
\end{figure}

We consider the transit probabilities of the four target planets with those of the full sample of RV-discovered exoplanets. We acquire the {\it prior} transit probabilities for 673 RV-discovered exoplanets from \citet{Dalba19}, which we display in Figure \ref{fig:trans_prob}. These probabilities do not account for the known mass distribution of the sample of RV exoplanets and, therefore, are not {\it posterior} transit probabilities \citep{Ho2011, Stevens2013}. The transit probabilities we calculated with Equation \ref{eq:transprob} for HD~9446~b~and~c,  HD~43691~b,  and HD~179079~b are also prior probabilities, meaning they can be compared with those from \citet{Dalba19}. All four of these transit probabilities fall within one standard deviation of the median of the distribution (Figure \ref{fig:trans_prob}). However, all probabilities except that of HD~9446~c are higher than the median. This suggests that these targets were relatively good candidates for transit searches especially considering their orbital periods. 



\section{Radial Velocity Analysis}
\label{RVA}

\subsection{Data Acquisition: Spectroscopy}

Spectroscopic data for each target was obtained by the high resolution echelle spectrometer (HIRES) on the 10.0-m Keck I telescope. The  details  of the  instrument  hardware  and  data  reduction  are  described  in  detail  by \citet{Vogt94}. This data, along with that available from the ELODIE \citep{Bara96} and SOPHIE spectrographs \citep{SOPHIE08} was used to confirm the orbital solutions of the systems using RadVel (see Table \ref{tab:Spectr} for observation details). The complete RV dataset analyzed is available in Appendix \ref{APP}.

\begin{deluxetable*}{llllll}
\tablecaption{Log of Spectroscopic Observations \label{tab:Spectr}}
\tablehead{
  \colhead{System} & 
  \colhead{Instrument} & 
  \colhead{Dates} & 
  \colhead{Passband} &
  \colhead{$N_{data}$}
}
\startdata
HD9446 & HIRES & 2013 Oct 23 -- 2019 Oct 18 & $300 - 1100$ nm & 51 \\
 & SOPHIE & 2006 Nov 3 -- 2009 Mar 3 & $387.2- 694.3$ nm & 79 \\
HD43691 & HIRES & 2004 Jan 10 -- 2019 Sep 6 & $300 - 1100$ nm & 31 \\
 & SOPHIE & 2006 Nov 5 -- 2007 Feb 23 & $387.2- 694.3$ nm & 14 \\
 & ELODIE & 2004 Nov 24 -- 2006 May 16 & $389.5 - 681.5$ nm & 22 \\
HD179079 & HIRES & 2004 Jul 11 -- 2019 Aug 28 & $300 - 1100$ nm & 93
\enddata
\end{deluxetable*}

\subsection{Orbital Solutions \& Linear Trends}
\label{OS}

Here we present the revised orbital solutions for HD~9446, HD~43691, and HD~179079.

Each star's RV data was fit using the RV modeling toolkit RadVel \citep{Fulton18} in order to confirm the orbital solution and look for linear trends to determine if there were indications for additional planetary companions. RadVel enables users to model Keplerian orbits in RV time series. RadVel fits RVs using maximum a posteriori probability (MAP) and employs ``modern Markov chain Monte Carlo (MCMC) sampling techniques and robust convergence criteria to ensure accurately estimated orbital parameters and their associated uncertainties" \citep{Fulton18}. Once the MCMC chains are well mixed and the orbital parameters which maximize the posterior probability are found, RadVel then supplies an output of the final parameter values from the MAP fit, provides RV time series plots and MCMC corner plots showing all joint posterior distributions derived from the MCMC sampling.

RadVel also provides a mid transit time of each planet along with uncertainties, which when combined with the expected transit duration of the planet defines the estimated transit window of the planet.

\begin{deluxetable}{lrrr}
\tablecaption{ MCMC Posteriors HD9446 \label{tab:params}}
\tablehead{
  \colhead{Parameter} & 
  \colhead{Credible Interval} & 
  \colhead{MAP} & 
  \colhead{Units}
}
\startdata
\sidehead{\bf{MCMC}}
  $P_{b}$ & $30.0608^{+0.0034}_{-0.0033}$ & $30.0607$ & days\\
  $T\rm{conj}_{b}$ & $2457790.96^{+0.56}_{-0.58}$ & $2457790.98$ & JD \\
  $e_{b}$ & $0.214^{+0.04}_{-0.041}$ & $0.22$ &  \\
  $\omega_{b}$ &  $-2.09^{+0.21}_{-0.22}$ & $-2.08$ & radians \\
  $K_{b}$ & $46.0\pm 2.1$ & $46.1$ & m s$^{-1}$ \\
  $P_{c}$ & $189.6\pm 0.13$ & $189.58$ & days \\
  $T\rm{conj}_{c}$ & $2457724.1\pm 2.6$ & $2457724.1$ & JD \\
  $e_{c}$ & $0.071^{+0.031}_{-0.032}$ & $0.067$ &  \\
  $\omega_{c}$ & $-2.29^{+5.2}_{-0.65}$ & $-3.0$ & radians \\
  $K_{c}$ & $60.8^{+1.9}_{-2.0}$ & $61$ & m s$^{-1}$ \\
\hline
\sidehead{\bf{Other}}
   $\gamma_{\rm SOPHIE}$ & $\equiv44.2309$ & $\equiv44.2309$ & m s$^{-1}$ \\
  $\gamma_{\rm HIRES}$ & $\equiv1.4458$ & $\equiv1.4458$ & m s$^{-1}$ \\
  $\dot{\gamma}$ & $0.0115^{+0.0025}_{-0.0023}$ & $0.0116$ & m s$^{-1}$d$^{-1}$ \\
  $\ddot{\gamma}$ & $\equiv0.0$ & $\equiv0.0$ & m s$^{-1}$d$^{-2}$ \\
  $\sigma_{\rm SOPHIE}$ & $-15.8^{+1.5}_{-1.7}$ & $-15.0$ & $\rm m\ s^{-1}$ \\
  $\sigma_{\rm HIRES}$ & $11.4^{+1.4}_{-1.2}$ & $10.5$ & $\rm m\ s^{-1}$ \\
\hline
\sidehead{\bf{Derived}}
    $M_b\sin i$  & $0.687^{+0.056}_{-0.055}$ & $0.709$ & M$_{\rm J}$ \\
  $a_b$  & $0.1892^{+0.0061}_{-0.0065}$ & $0.1844$ & AU \\
  $M_c\sin i$  & $1.71\pm 0.13$ & $1.66$ & M$_{\rm J}$ \\
  $a_c$  & $0.646^{+0.021}_{-0.022}$ & $0.629$ & AU \\
\enddata
\tablenotetext{}{
  Reference epoch for $\gamma$,$\dot{\gamma}$,$\ddot{\gamma}$: 2457752.165903 \\ NOTE: Negative jitter values in the RadVel results are equal to the positive equivalent. 
}
\end{deluxetable}


\begin{figure}[tbh]
\centering 
\includegraphics[width=0.48\textwidth]{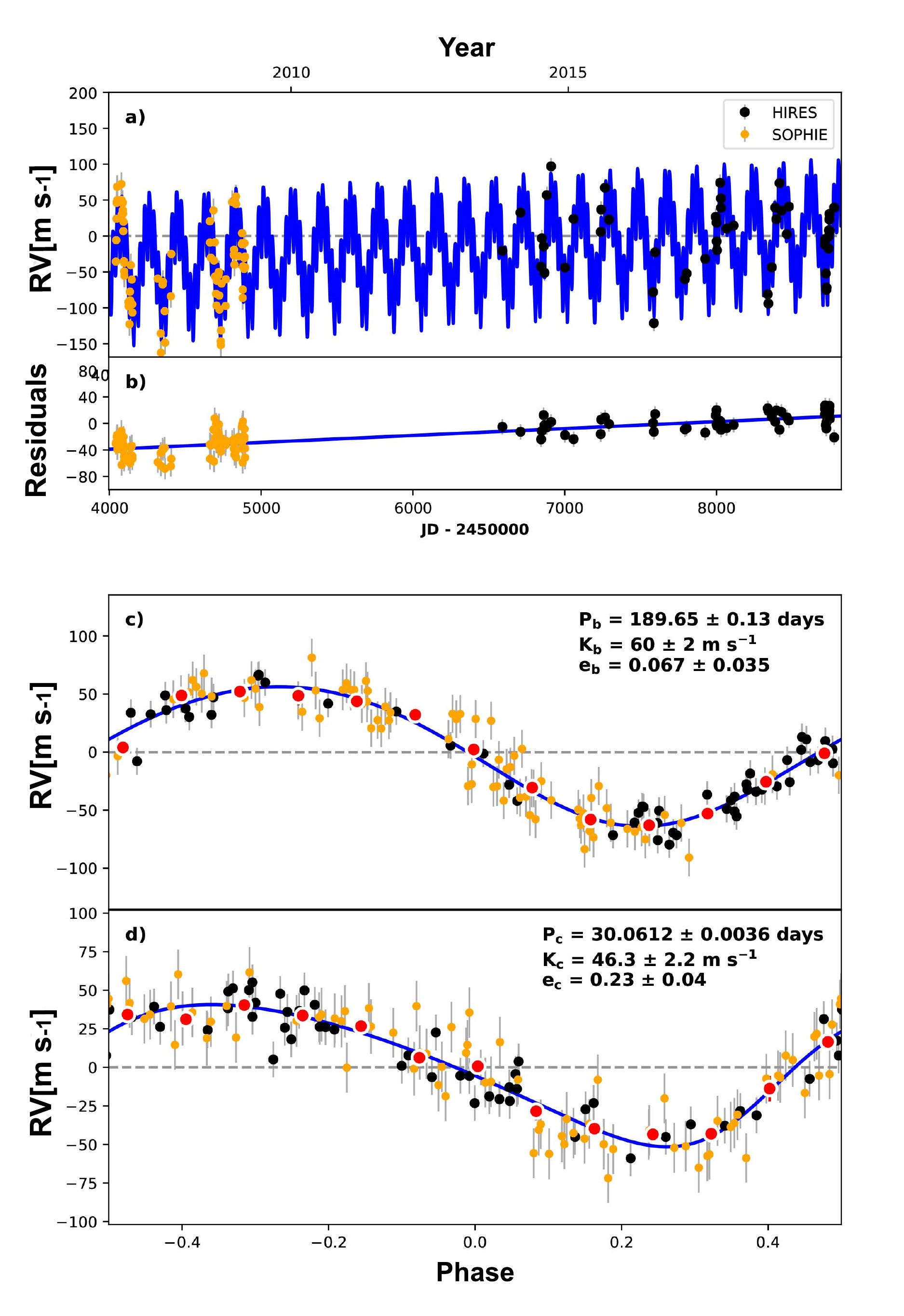}
\caption{ Best-fit 2-planet Keplerian orbital model for HD9446. The maximum likelihood model is plotted while the orbital parameters listed in Table \ref{tab:params} are the median values of the posterior distributions.  The thin blue line is the best-fit 2-planet model. We add in quadrature the RV jitter terms listed in Table \ref{tab:params} with the measurement uncertainties for all RVs.  {\bf b)} Residuals to the best-fit 2-planet model. {\bf c) \& d)} RVs phase-folded to the ephemeris of planets b \& c respectively.  The small point colors and symbols are the same as in panel {\bf a}.  Red circles are the same velocities binned in 0.08 units of orbital phase.  The phase-folded model for planets b \& c are shown as the blue lines.}
\label{fig:HD9446RV}
\end{figure}

\begin{figure}[tbh]
\centering 
\includegraphics[width=0.5\textwidth]{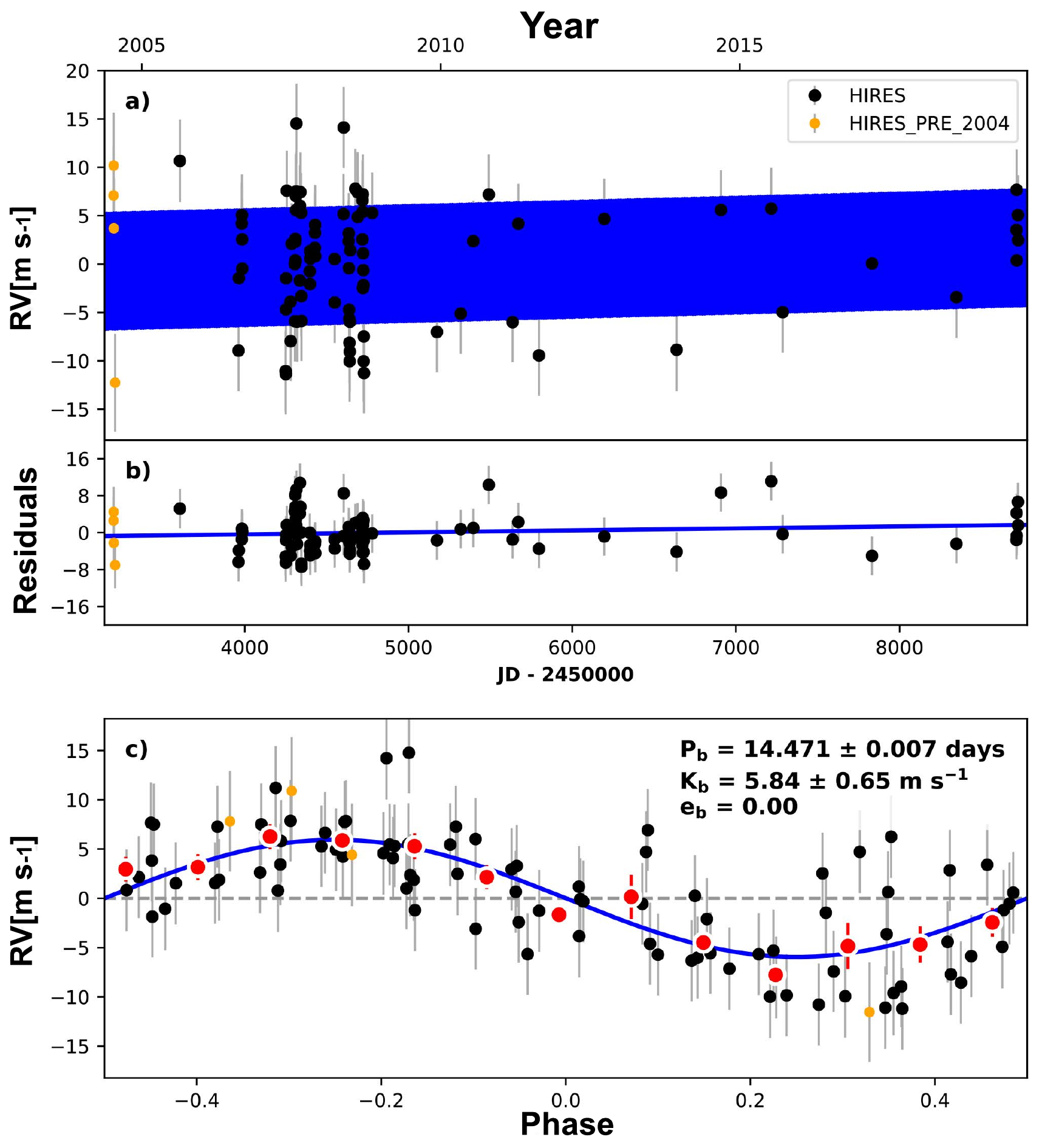}
\caption{ Best-fit 1-planet Keplerian orbital model
  for HD~179079. The maximum likelihood model is plotted while
  the orbital parameters listed in Table \ref{tab:params2} are the
  median values of the posterior distributions.  The thin blue line is
  the best-fit 1-planet model. We add in quadrature
  the RV jitter terms listed in Table \ref{tab:params2} with the
  measurement uncertainties for all RVs.  {\bf b)} Residuals to the
  best-fit 1-planet model. {\bf c)} RVs phase-folded
  to the ephemeris of planet b.  The small point colors
  and symbols are the same as in panel {\bf a}.  Red circles are the same velocities binned in 0.08 units of orbital
  phase.  The phase-folded model for planet b is shown as the blue
  line.}
\label{fig:HD179079RV}
\end{figure}

\begin{figure}[tbh]
\centering 
\includegraphics[width=0.5\textwidth]{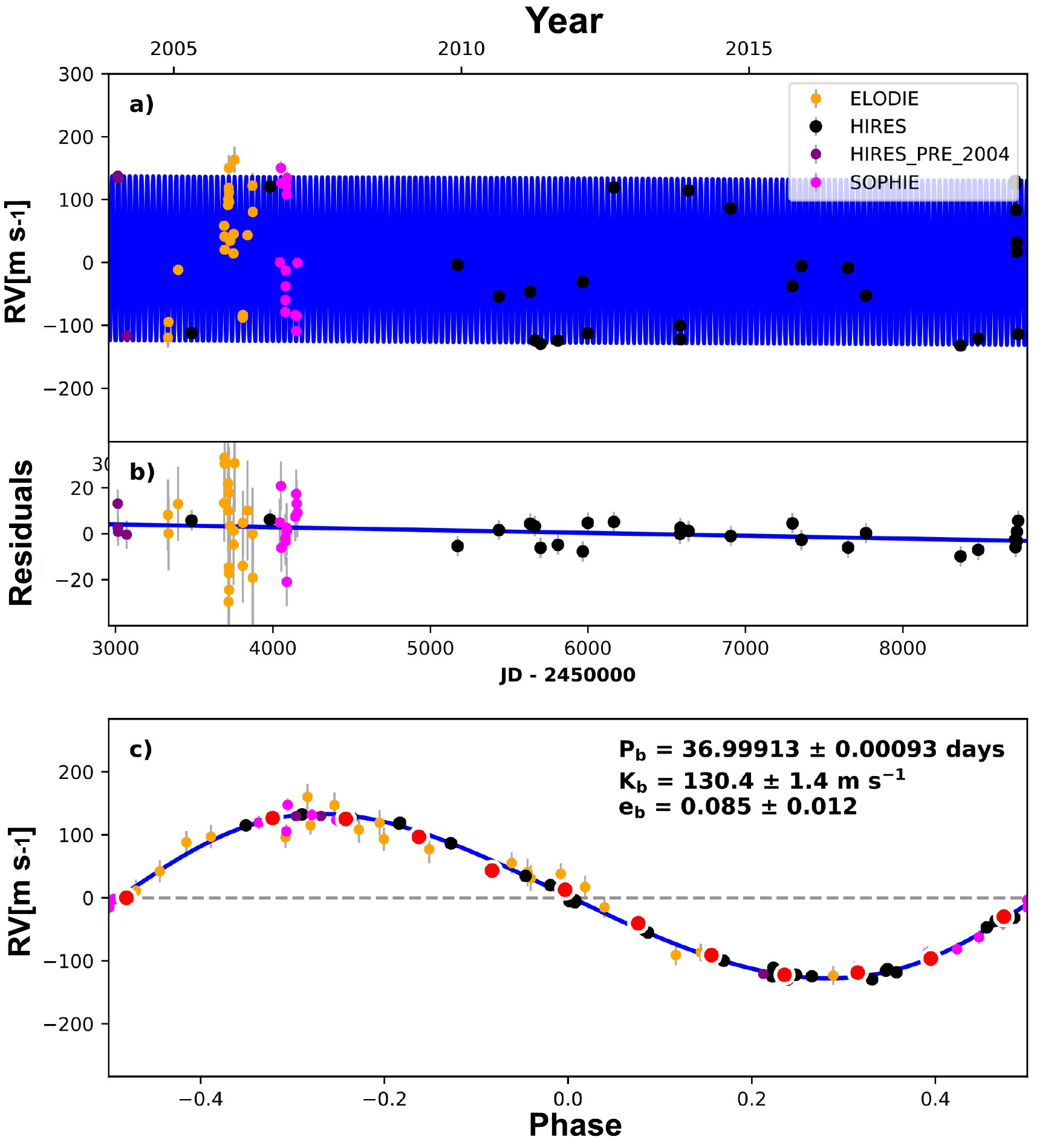}
\caption{ Best-fit 1-planet Keplerian orbital model
  for HD~43691. The maximum likelihood model is plotted while
  the orbital parameters listed in Table \ref{tab:params3} are the
  median values of the posterior distributions.  The thin blue line is
  the best-fit 1-planet model. We add in quadrature
  the RV jitter terms listed in Table \ref{tab:params3} with the
  measurement uncertainties for all RVs.  {\bf b)} Residuals to the
  best-fit 1-planet model. {\bf c)} RVs phase-folded
  to the ephemeris of planet b. The small point colors
  and symbols are the same as in panel {\bf a}.  Red circles are the same velocities binned in 0.08 units of orbital
  phase.  The phase-folded model for planet b is shown as the blue
  line.}
\label{fig:HD43691RV}
\end{figure}

A Keplerian orbital solution was fit to the RV data for HD~9446, HD~179079 and HD~43691 using RadVel in order to refine the orbital parameters and transit ephemeris of each system. Public RV data from SOPHIE and ELODIE published in \citet{Dasilva07} and \citet{Hebrard10} was combined with our Keck HIRES data which is provided in full in Appendix \ref{APP}. Gamma velocities of 21712.56 $m s^{-1}$, -28967.79$m s^{-1}$ and -28958.18$m s^{-1}$ were subtracted from the total RV values of the SOPHIE data for HD~9446, SOPHIE data for HD~43691 and ELODIE data for HD~43691 respectively before being run through RadVel. Thus the gamma values in Tables \ref{tab:params}, \ref{tab:params2} and \ref{tab:params3} ($\sigma$) are in reference to these values. For Tables \ref{tab:params}, \ref{tab:params2} and \ref{tab:params3} list the resulting median values of the posterior distributions for each parameter in the orbital solutions. The best-fit Keplerian orbital model for the system and planet orbital solutions are shown for the HD~9446~b and HD~9446~c planets in Figure \ref{fig:HD9446RV},  for planet HD~179079~b in Figure \ref{fig:HD179079RV}, and for planet HD~43691~b in Figure \ref{fig:HD43691RV}. Comparing these results with those from Table \ref{tab:Stellar}, each planet's orbital solution is in close agreement with the published values with the exception of HD~9446~c. Our fit presents a new best-fit period of this planet of 189.6$\pm 0.13$ days. Note all other variances for HD~9446~c fall within the uncertainties of the published values.
We fit each system allowing the trend parameter to be free in order to look for linear trends. These trends could be indications of additional unknown planets in orbit around the host star. RadVel provides a model comparison summary with both Akaike information criterion (AIC) and Bayesian information criterion (BIC) values. The results from these model comparisons can be found in Tables \ref{tab:BIC_9446}, \ref{tab:BIC_179079} $\&$ \ref{tab:BIC_43691} in Appendix \ref{MCT}. For HD~179079 both AIC and BIC methods preferred the model without a trend (See Table \ref{tab:BIC_179079}). For HD~43691 both the AIC and BIC preferred the model with a $\sim1.75\sigma$ trend (See Tables \ref{tab:params3} $\&$ \ref{tab:BIC_43691}). While this could indicate the presence of another unknown planet in the system, this is a low sigma result and so warrants only a small mention here. For HD~9446 both comparison methods prefer models with the trend parameter free, with the best fit including a $\sim4.8\sigma$ trend (See Tables \ref{tab:params} $\&$ \ref{tab:BIC_9446}). As the results from our speckle imaging analysis in Section \ref{SP} indicated no evidence of a stellar companion, this significant trend likely indicates an additional planet in the HD~9446 system. Further RV observation of this system is required to determine the cause of this linear trend. Note the AIC and BIC comparisons of HD~9446 preferred different models (BIC preferred $e_b$ as a fixed parameter). However, for the purposes of determining looking for linear trends, each comparison method preferred a model with a trend. 

As well as providing updated values for each planet's period, eccentricity and semi-amplitude, RadVel also derives the values for the planet semi-major axis (a) and mass ($M_p \sin i$). These values are shown at the bottom of Tables \ref{tab:params}, \ref{tab:params2} and \ref{tab:params3}. 

Using the posteriors for the time of conjunction from the MCMC output, we calculated the transit window for each planet to use in determining if a transit of these planets had been detected (See Section~\ref{PV}).

\begin{deluxetable}{lrrr}
\tablecaption{ MCMC Posteriors HD~179079 \label{tab:params2}}
\tablehead{
  \colhead{Parameter} & 
  \colhead{Credible Interval} & 
  \colhead{MAP} & 
  \colhead{Units}
}
\startdata
\sidehead{\bf{MCMC}}
$P_{b}$ & $14.4708^{+0.01}_{-0.0035}$ & $14.47$ & days \\
  $T\rm{conj}_{b}$ & $2454881.84^{+0.33}_{-0.29}$ & $2454881.73$ & JD \\
  $e_{b}$ & $\equiv0.0$ & $\equiv0.0$ &  \\
  $\omega_{b}$ & $\equiv-0.0$ & $\equiv-0.0$ & radians \\
  $K_{b}$ & $5.84^{+0.63}_{-0.66}$ & $5.95$ & m s$^{-1}$ \\
\hline
\sidehead{\bf{Other}}
 $\gamma_{\rm HIRES_{PRE 2004}}$ & $\equiv-2.2151$ & $\equiv-2.2151$ & m s$^{-1}$ \\
  $\gamma_{\rm HIRES}$ & $\equiv-2.0527$ & $\equiv-2.0527$ & m s$^{-1}$ \\
  $\dot{\gamma}$ & $0.00048^{+0.00039}_{-0.00037}$ & $0.00043$ & m s$^{-1}$d$^{-1}$ \\
  $\ddot{\gamma}$ & $\equiv0.0$ & $\equiv0.0$ & m s$^{-1}$d$^{-2}$ \\
  $\sigma_{\rm HIRES_{PRE 2004}}$ & $7.4^{+7.9}_{-3.5}$ & $4.6$ & $\rm m\ s^{-1}$ \\
  $\sigma_{\rm HIRES}$ & $4.13^{+0.35}_{-0.33}$ & $3.97$ & $\rm m\ s^{-1}$ \\
\hline
\sidehead{\bf{Derived}}
 $a_b$  & $0.1214^{+0.0064}_{-0.0071}$ & $0.1014$ & AU \\
  $M_b\sin i$  & $0.076\pm{0.012}$ & $0.053$ & M$_{\rm J}$ \\
\enddata
\tablenotetext{}{
  Reference epoch for $\gamma$,$\dot{\gamma}$,$\ddot{\gamma}$: 2454872.463263
}
\end{deluxetable}

 
\begin{deluxetable}{lrrr}
\tablecaption{ MCMC Posteriors HD43691 {\label{tab:params3}} }
\tablehead{
  \colhead{Parameter} & 
  \colhead{Credible Interval} & 
  \colhead{MAP} & 
  \colhead{Units}
}
\startdata
\sidehead{\bf{MCMC}}
$P_{b}$ & $36.99913^{+0.00095}_{-0.00092}$ & $36.99932$ & days \\
  $T\rm{conj}_{b}$ & $2456282.91^{+0.15}_{-0.14}$ & $2456282.89$ & JD \\
  $e_{b}$ & $0.085^{+0.012}_{-0.011}$ & $0.087$ &  \\
  $\omega_{b}$ & $-1.35\pm 0.12$ & $-1.34$ & radians \\
  $K_{b}$ & $130.4\pm 1.4$ & $130.6$ & $\rm m\ s^{-1}$ \\
\hline
\sidehead{\bf{Other}}
$\gamma_{\rm SOPHIE}$ & $\equiv-12.4134$ & $\equiv-12.4134$ & m s$^{-1}$ \\
  $\gamma_{\rm HIRES_{PRE 2004}}$ & $\equiv21.0278$ & $\equiv21.0278$ & m s$^{-1}$ \\
  $\gamma_{\rm HIRES}$ & $\equiv21.6117$ & $\equiv21.6117$ & m s$^{-1}$ \\
  $\gamma_{\rm ELODIE}$ & $\equiv-45.0874$ & $\equiv-45.0874$ & m s$^{-1}$ \\
  $\dot{\gamma}$ & $-0.00121^{+0.00069}_{-0.00068}$ & $-0.00124$ & m s$^{-1}$d$^{-1}$ \\
  $\ddot{\gamma}$ & $\equiv0.0$ & $\equiv0.0$ & m s$^{-1}$d$^{-2}$ \\
  $\sigma_{\rm SOPHIE}$ & $10.7^{+3.3}_{-2.3}$ & $9.8$ & $\rm m\ s^{-1}$ \\
  $\sigma_{\rm HIRES_{PRE 2004}}$ & $-3.2^{+16.0}_{-9.6}$ & $-6$ & $\rm m\ s^{-1}$ \\
  $\sigma_{\rm HIRES}$ & $4.85^{+0.95}_{-0.74}$ & $4.16$ & $\rm m\ s^{-1}$ \\
  $\sigma_{\rm ELODIE}$ & $-3^{+16}_{-10}$ & $-11$ & $\rm m\ s^{-1}$ \\
\hline
\sidehead{\bf{Derived}}  
  $a_b$  & $0.238^{+0.014}_{-0.016}$ & $0.227$ & AU \\
  $M_b\sin i$  & $2.57^{+0.31}_{-0.34}$ & $2$ & M$_{\rm J}$ \\
\enddata
\tablenotetext{}{
  Reference epoch for $\gamma$,$\dot{\gamma}$,$\ddot{\gamma}$: 2456284.051544 \\
  NOTE: Negative jitter values in the RadVel results are equal to the positive equivalent. }
\end{deluxetable}

 
\subsection{Search for Additional Planets and Injection-Recovery Tests}
\label{IT}

We searched for additional planet candidates in our radial velocity data using RVSearch (Rosenthal et al. in prep), an iterative periodogram algorithm. First we define an orbital frequency/period grid over which to search, with sampling such that the difference in frequency between adjacent grid points is $\frac{1}{2\pi \tau}$, where $\tau$ is the observational baseline. Using this grid we compute a goodness-of-fit periodogram by fitting a sinusoid with a fixed period to the data for each period in the grid. We choose to measure goodness-of-fit as the change in the Bayesian Information Criterion (BIC) at each grid point between the best-fit 1-planet model with the given fixed period, and the BIC value of the 0-planet fit to the data. We then fit a power law to the noise histogram (50-95 percent) of the data and accordingly extrapolate a BIC detection threshold corresponding to an empirical false-alarm probability of our choice (we choose 0.003). 

If one planet is detected we perform a final fit to the one-planet model with all parameters free, including eccentricity, and record the BIC of that best-fit model. We then add a second planet to our RV model and conduct another grid search, leaving the parameters of the first planet free to converge to a more optimal solution. In this case we compute the goodness-of-fit as the difference between the BIC of the best-fit one-planet model, and the BIC of the two-planet model at each fixed period in the grid. We set a detection threshold in the manner described above and continue this iterative search until the n+1th search rules out additional signals. Once we have a complete orbital model we can characterize the completeness of our search using injection-recovery tests, wherein we inject synthetic planet signals into the data and check whether RVSearch can recover that synthetic signal.

\begin{figure}
\centering 
\subfloat{%
  \includegraphics[width=1.0\columnwidth]{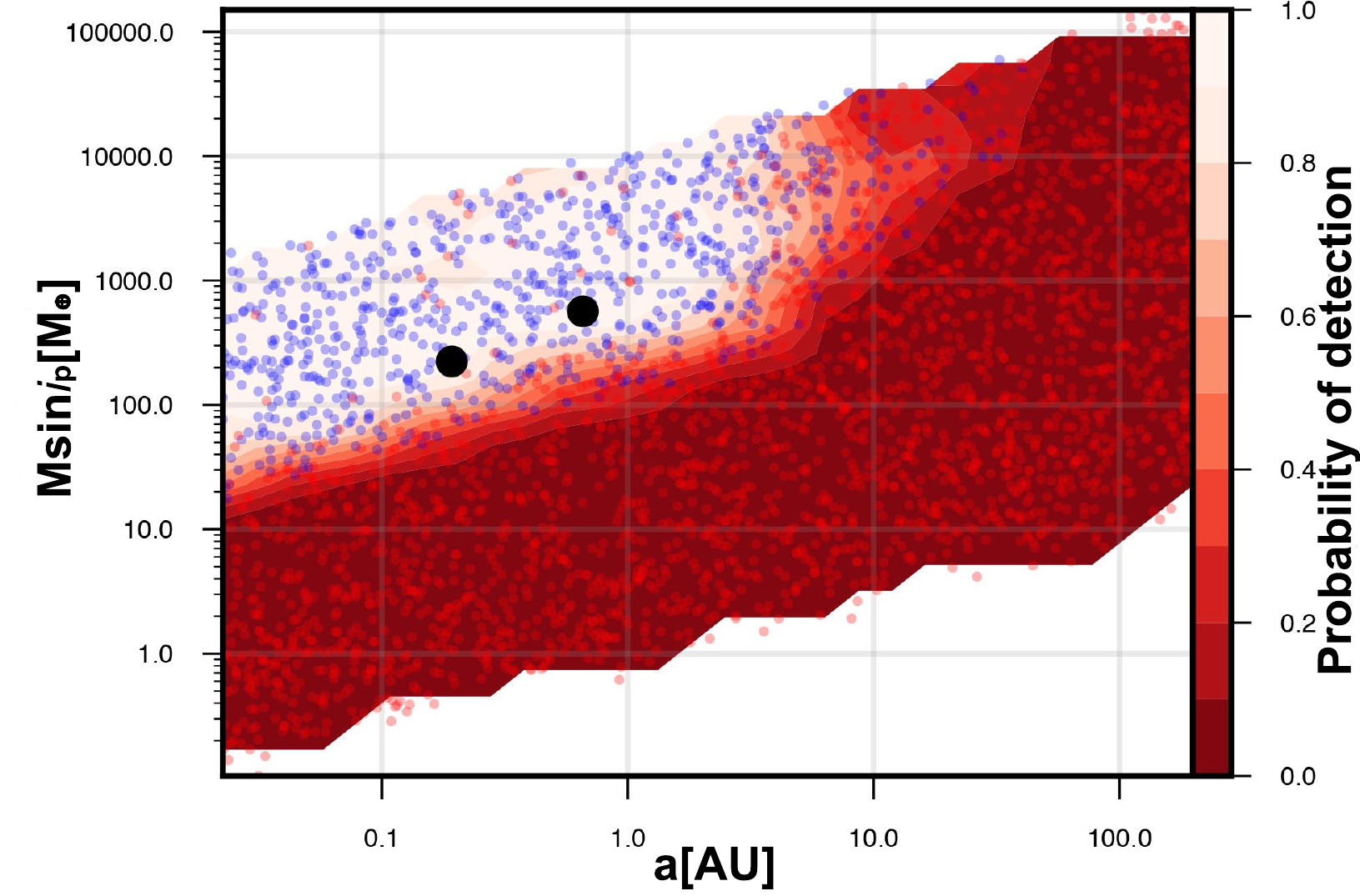}%
}\qquad
\subfloat{%
  \includegraphics[width=1.0\columnwidth]{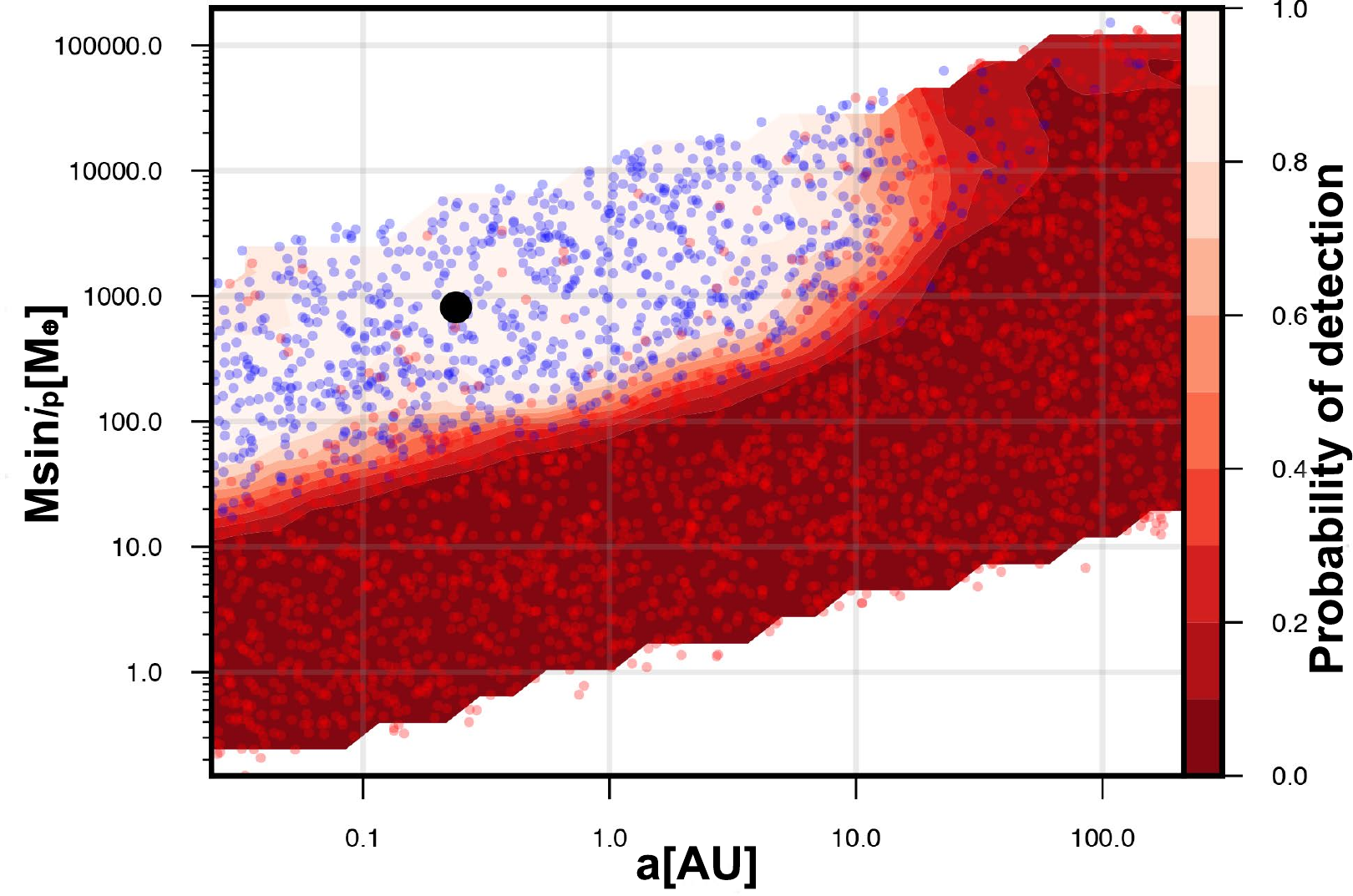}%
}\qquad
\subfloat{%
  \includegraphics[width=1.0\columnwidth]{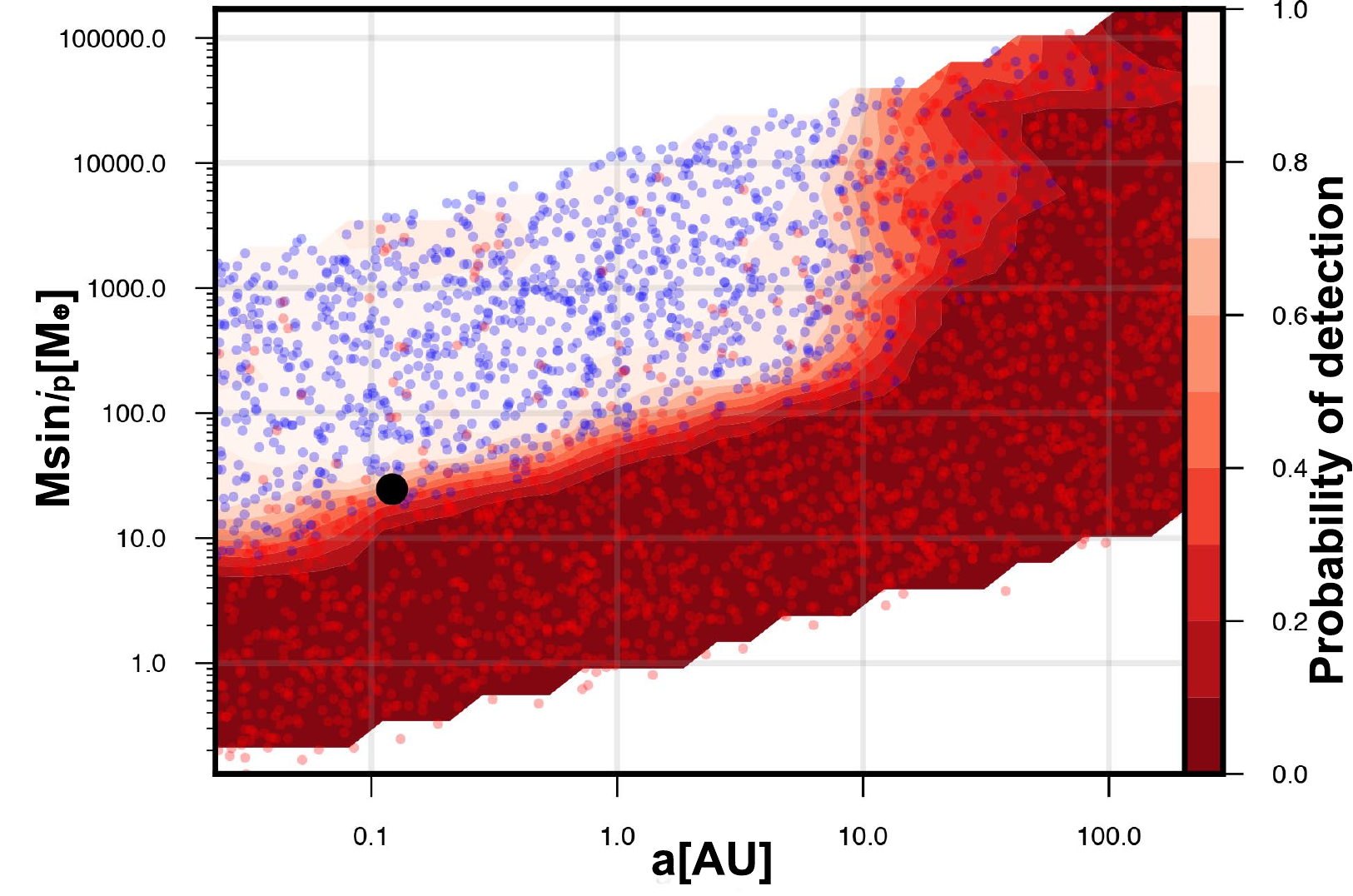}%
}
\caption{Injection recoveries for Top: HD~9446, Middle: HD~43691, and Bottom: HD~179079. Black dots indicate the known planets for each system, purple dots are injected planets that were successfully recovered and red dots are injected planets that were not recovered. The scale on the right of each plot shows the probability of recovery indicated by the contours on each plot. HD~9446~b~\&~c and HD~43691~b are each in a parameter space with a high probability of recovery whereas HD~179079~b borders the area of low probability. \label{fig:inject}}
\end{figure}

RVSearch was run on HD~9446, HD~43691 and HD~179079, and it successfully recovered each of the known planets in these systems. No additional planets were found via the search. Injection-recovery tests were then completed for each system to determine the completeness of the search, the results of which can be seen in Figure \ref{fig:inject}. Figure \ref{fig:inject} shows the detection results of the injection-recovery tests for HD~9446 (top), HD~43691 (middle), and HD~179079 (bottom). The black dots indicate the known planets for each system whereas the smaller dots represent the injected planets. Dots that are purple are injected planets that were successfully recovered, and dots that are red are planets that were not recovered. The contouring shows the parameter spaces in which planets are likely to be recovered in light pink and those that have a low probability of detection in dark red. HD~9446~b~\&~c and HD~43691~b are each in a parameter space with a high probability of recovery whereas HD~179079~b borders the area of low probability.

Following the methods outlined in \citet{Brandt_2019} and \citet{Kiefer2019} we then attempted to constrain the mass and period of the additional companion in the HD~9446 system by calculating the difference between the Gaia \citep{Gaia18} and Hipparcos \citep{Vanleeuwen2007} positions for HD~9446, deriving the proper motion and comparing the value with the Gaia proper motion. We found no evidence from the Hipparcos to Gaia positional difference to indicate there is an acceleration. The derived proper motion from the difference between the Hipparcos and Gaia positions is consistant to the Gaia proper motion to within 1 sigma, and so we did not attempt further to constrain the period and mass of the additional companion.

\section{Photometry Analysis}
\label{PV}

\subsection{Data Acquisition: KELT Photometry}

The Kilodegree Extremely Little Telescope (KELT) survey uses two robotic, wide-field telescopes to photometrically survey stars in the magnitude range 8~$<$~V~$<$~11 for transiting planets \citep{Pepper2007,Pepper2012}.  It observes a set of predefined fields of size $26^\circ \times 26^\circ$, about once every 20 minutes, as long as the fields are visible.  KELT uses a nonstandard passband, which is similar to a broad $R$-band. Details of the exoplanet discoveries and additional science results from the KELT survey are described in \citet{Pepper2018}. 

KELT observed HD 9446 in survey field KN02 8467 times from 2006 October 26 through 2014 December 27.  The light curve has an RMS scatter of 4.7 mmag.  It observed HD 43691 in survey field KN04 9560 times from 2006 October 27 through 2014 December 30, with an RMS scatter of 8.4 mmag.  It observed HD 179079 in both survey fields KS13 and KS14.  KELT observed HD 179079 in field KS13 5835 times from 2010 March 20 through 2017 September 23, with an RMS scatter of 6.3 mmag, and in field KS14 3276 times from 2010 April 12 to 2014 October 23, with an RMS scatter of 7.2 mmag. These observational details are provided in Table \ref{tab:Photometric}.

\subsection{Data Acquisition: APT Photometry}

We observed HD~9446, HD~43691, and HD~179079 throughout multiple observing
seasons with the T10 and T12 0.80-m automatic photoelectric telescopes (APTs) 
at Fairborn Observatory in the Patagonia Mountains of southern Arizona. 
These APTs are two of several automated photometric, spectroscopic, and 
imaging telescopes operated by Tennessee State University at Fairborn. Both 
T10 and T12 are equipped with a two-channel precision photometer that uses a 
dichroic filter and two EMI 9124QB bi-alkali photomultiplier tubes to separate 
and simultaneously measure the Str\"omgren $b$ and $y$ passbands. The APTs are programmed to make differential brightness measurements of the program stars 
with respect to three comparison stars in the same field.  From the raw 
counts, the differential magnitudes are computed as the difference in 
brightness between the program star and the mean brightness of the two best 
comparison stars. The differential magnitudes are corrected for atmospheric 
extinction and transformed to the Str\"omgren system.  To improve the 
photometric precision, the differential $b$ and $y$ observations 
are combined into a single $(b+y)/2$ passband. Further details of the automatic telescopes,
precision photometers, and observing and data reduction procedures can be 
found in \citet{h1999} and \citet{ehf2003}.  Note that the T10 and T12 APTs 
are functionally identical to the T8 0.80-m APT described in \citet{h1999}.
More details on the analysis of the photometric observations for the TERMS
project can be found in \citet{hkw+2013}.

With the T10 APT, we obtained 318 observations of HD~9446 between 2015 September 27 and 2018 February 12 and 412 observations of HD~43691 between 2015 September 27 and 2018 March 15.  With the T12 APT, we obtained 703 observations of HD~179079 between 2007 June 6 and 2018 June 22.  The precision of the individual differential magnitudes is typically between 0.001 and 0.002 mag on good photometric nights. A table of these observational details is provided in Table~\ref{tab:Photometric}.

\begin{deluxetable*}{llllll}
\tablecaption{Log of Photometric Observations \label{tab:Photometric}}
\tablehead{
  \colhead{System} & 
  \colhead{Instrument} & 
  \colhead{Dates} & 
  \colhead{Passband} &
  \colhead{$N_{data}$}
}
\startdata
HD9446 & APT & 2015 Sep 27 -- 2018 Feb 12 & $b + y$  & 318 \\
 & KELT & 2006 Oct 26 -- 2014 Dec 27 & $R_{KELT}$  & 8467 \\
HD43691 & APT & 2015 Sep 27 -- 2018 Mar 15 & $b + y$  & 412 \\
 & KELT & 2006 Oct 27 -- 2014 Dec 30 & $R_{KELT}$  & 9560 \\
HD179079 & APT & 2007 Jun 6 -- 2018 Jun 22 & $b + y$  & 703 \\
 & KELT & 2010 Mar 20 -- 2017 Sep 23 & $R_{KELT}$  & 9111
\enddata
\end{deluxetable*}

\subsection{Search for Transits}
\label{SfT}

Using the refined transit ephemerides and orbital periods obtained in Section~\ref{OS} along with the photometric data from KELT and the T10 and T12 APTs, transits of the known planets were searched for in each system. 

\begin{figure*}
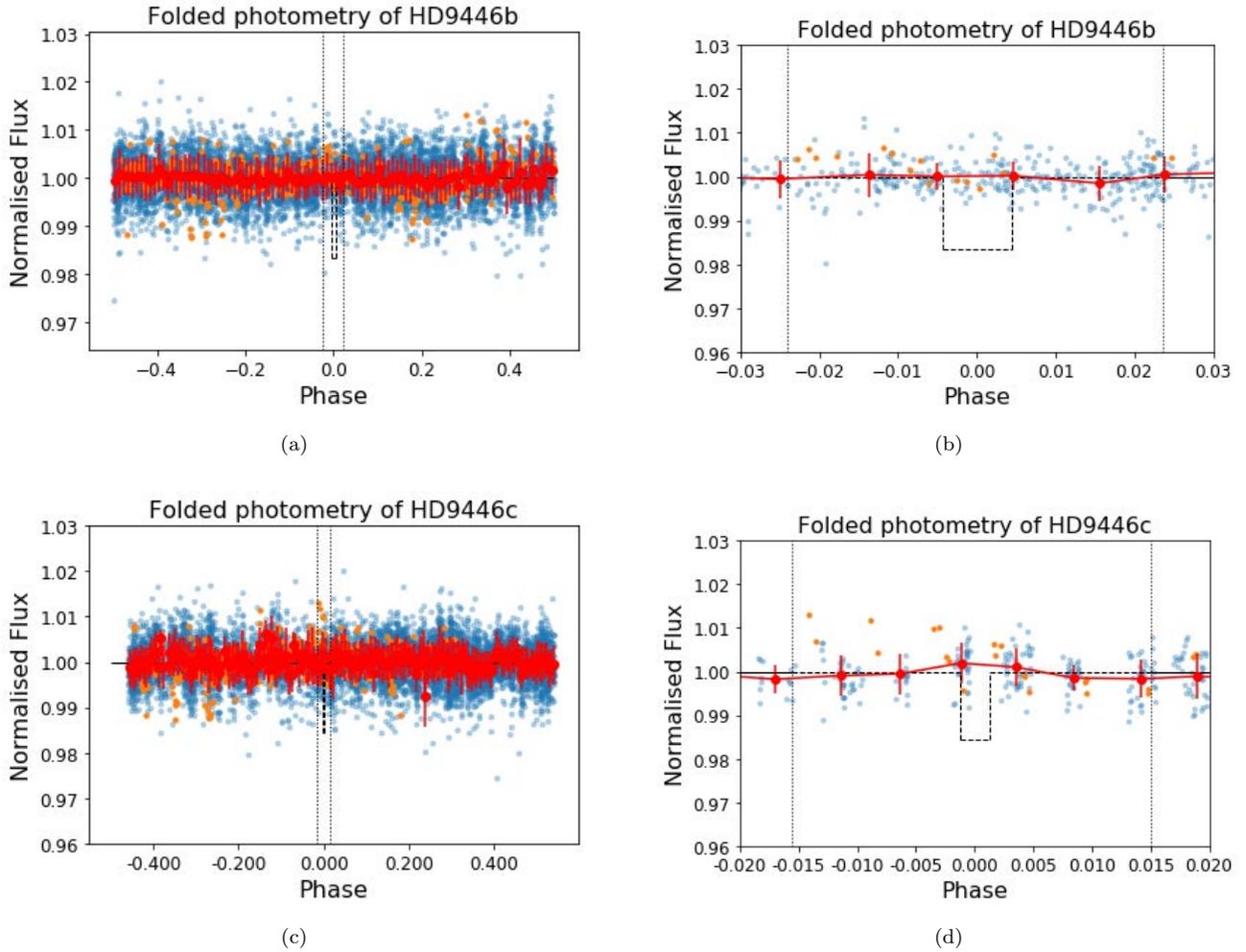

\gridline{\fig{"Folded_photometry_HD9446b"}{0.45\textwidth}{(a)}
          \fig{"Folded_photometry_HD9446b_zoom"}{0.45\textwidth}{(b)}}
\gridline{\fig{"Folded_photometry_HD9446c"}{0.45\textwidth}{(c)}
        \fig{"Folded_photometry_HD9446c_zoom"}{0.45\textwidth}{(d)}}
\caption{Top: (a) HD~9446~b phase folded KELT and APT photometry. (b) The same HD~9446~b data zoomed in on the transit window. 
Bottom: (c) HD~9446~c phase folded KELT and APT photometry. (d) The same HD~9446~c data zoomed in on the transit window. The transit window is defined by dotted vertical lines. The blue dots represent KELT data and orange dots represent APT data. The solid black line shows the transit baseline and the red line shows the binned average flux measurements for each planet. A dashed line shows the predicted transit signal expected for a central transit of the known planet. There is no significant difference between the normalized flux within the transit window compared to outside the transit window for either planet. For HD~9446~b this rules out a transit with a confidence of 3.96$\sigma$. Due to the data gap in the photometry of HD~9446~c we can rule out transits to a certainty of only 3.88 sigma for a planet with an impact parameter of up to b~=~0.5778.}
\label{fig:HD9446_phase}
\end{figure*}

\begin{figure*}
\gridline{\fig{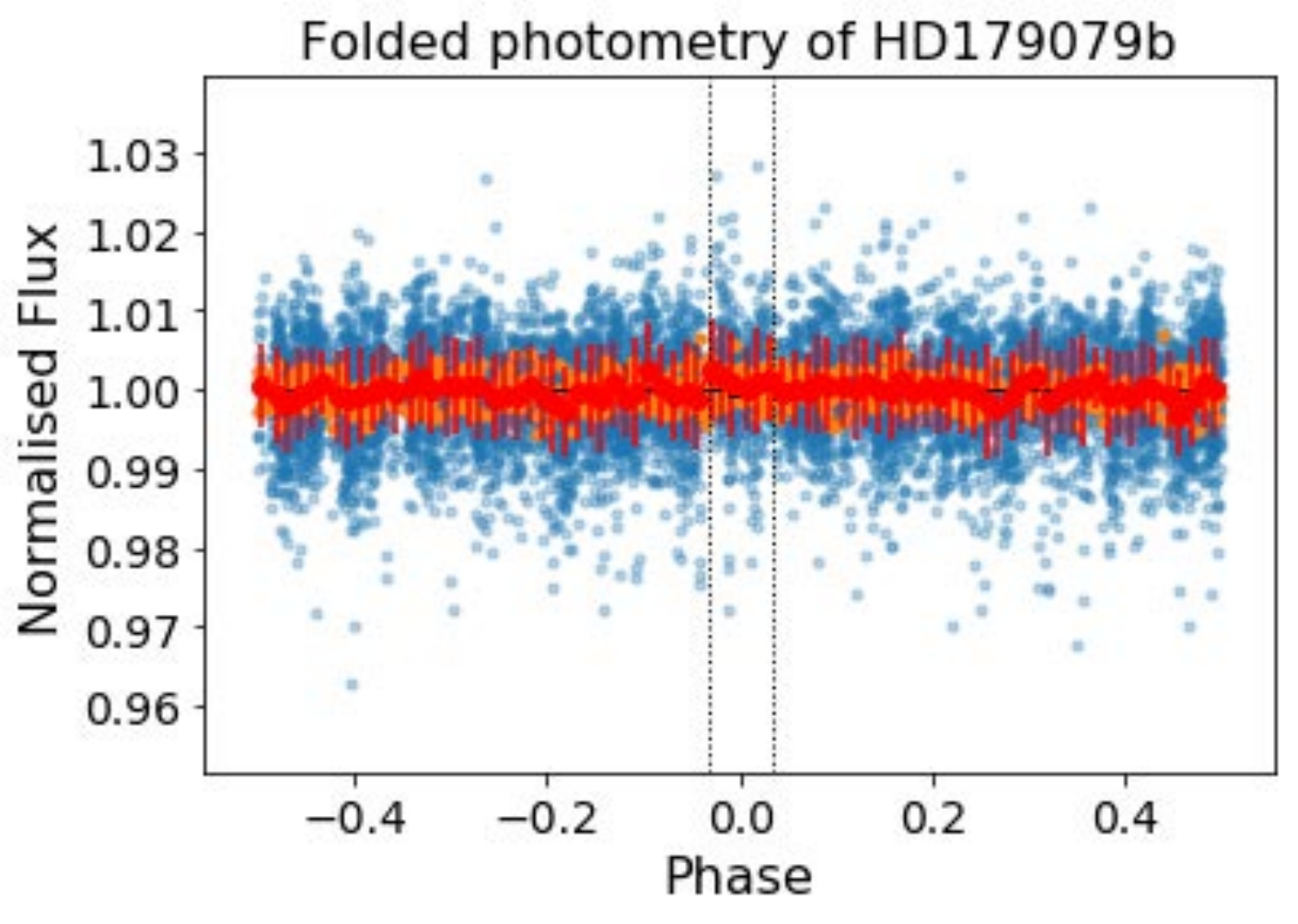}{0.45\textwidth}{(a)}
          \fig{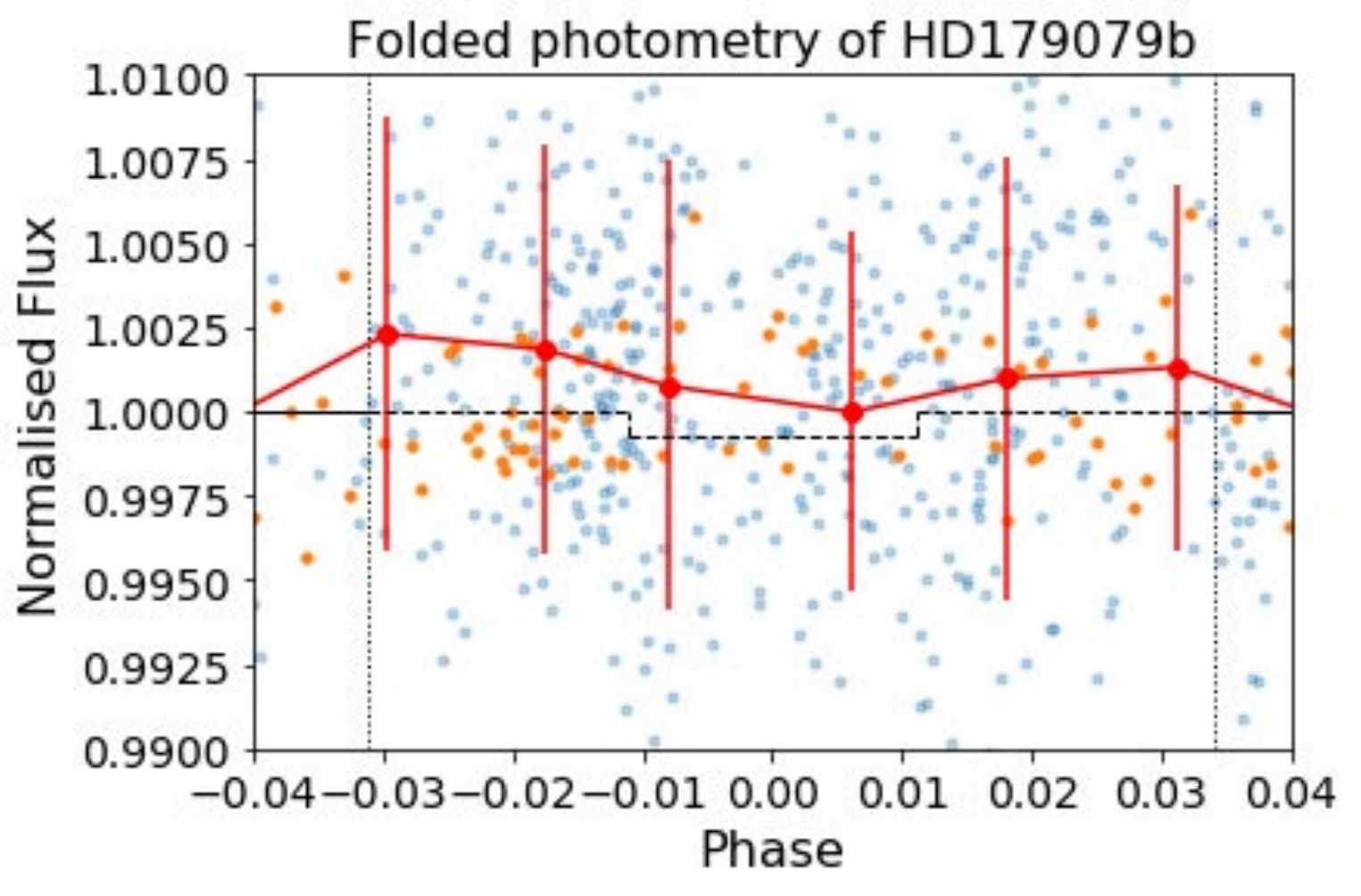}{0.45\textwidth}{(b)}}
\gridline{\fig{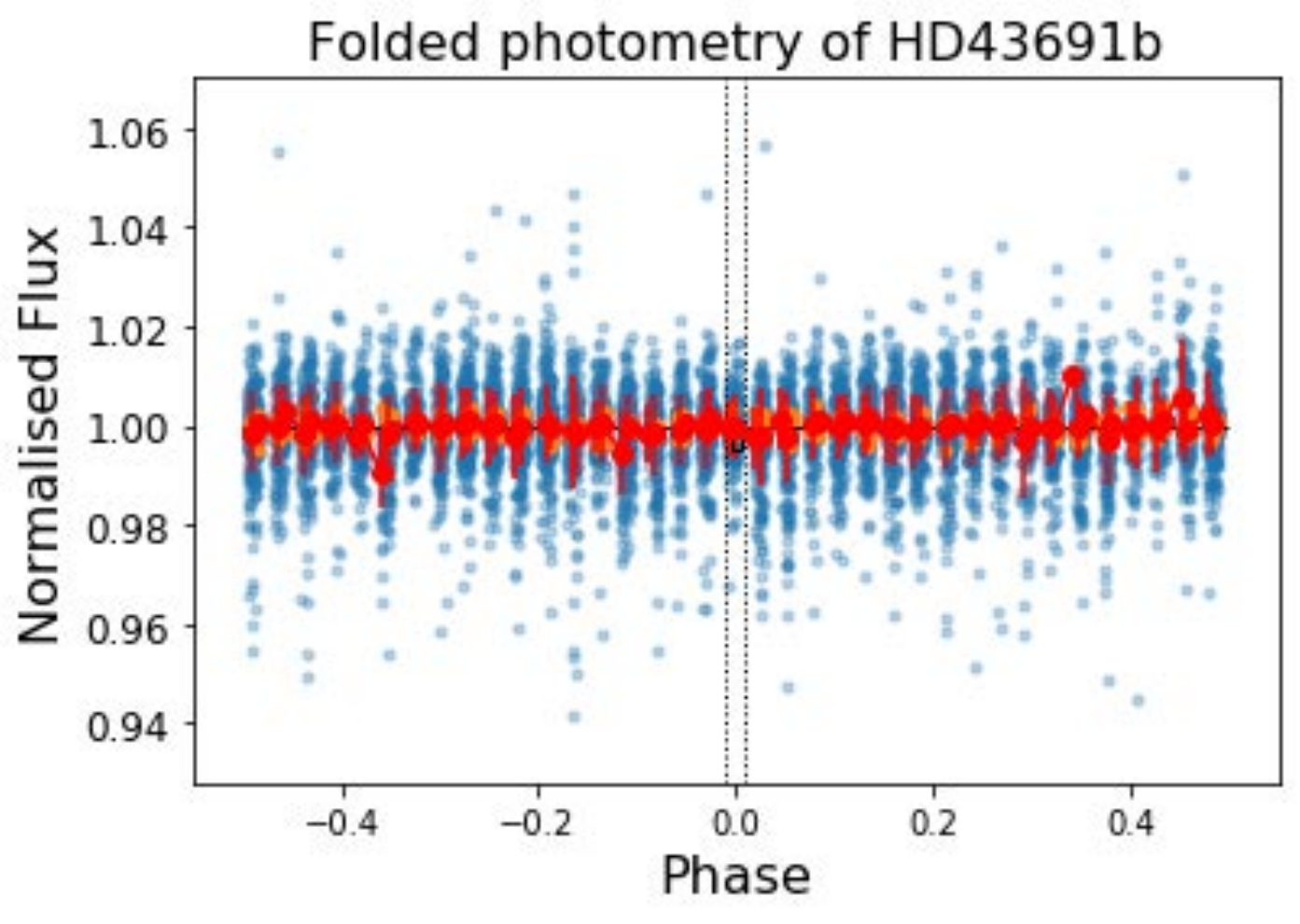}{0.45\textwidth}{(c)}
          \fig{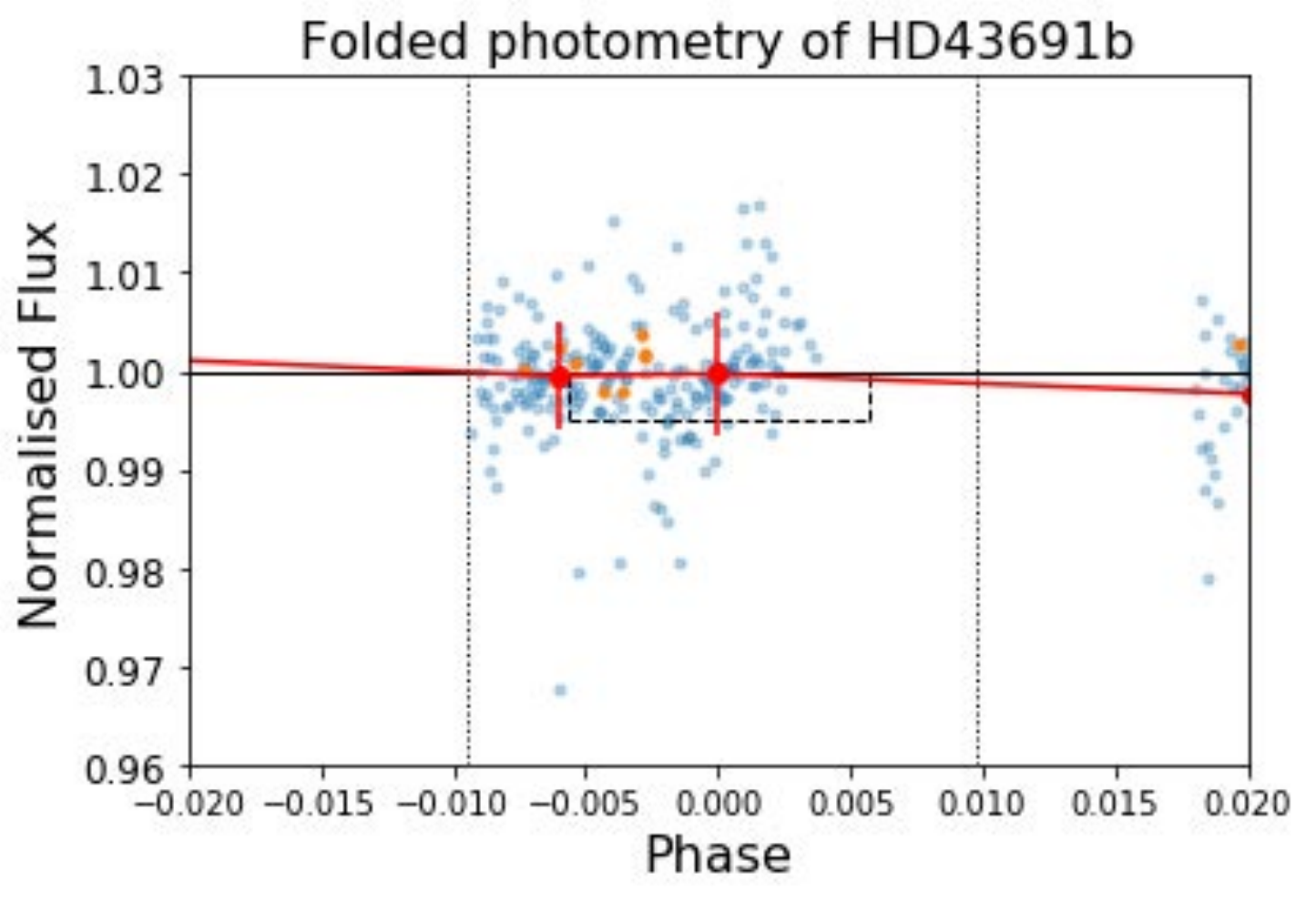}{0.45\textwidth}{(d)}}
\caption{Top: (a) HD~179079~b phase folded KELT and APT photometry. (b) The same HD~179079~b data zoomed in on the transit window. 
Bottom: (c) HD~43691~b phase folded KELT and APT photometry. (d) The same HD~43691~b data zoomed in on the transit window. The transit window is defined by dotted vertical lines. The blue dots represent KELT data and orange dots represent APT data. The solid black line shows the transit baseline and the red line shows the binned average flux measurements for each planet. A dashed line shows the predicted transit signal expected for a central transit of the known planet. There is no significant difference between the normalized flux within the transit window compared to outside the transit window for either system. Due to the data gap in the photometry of HD~43691 we can rule out transits to a certainty of only 0.74 sigma for a planet with an impact parameter of up to b~=~0.898. For HD~179079~b a transit cannot be ruled out due to the low signal to noise.}
\label{fig:HD43691b_phase}
\end{figure*}

The normalized observations from all observing seasons
are plotted in the top panels of Figures \ref{fig:HD9446_phase} and \ref{fig:HD43691b_phase} for HD~9446~b~and~c, HD~179079~b and HD~43691~b respectively.

The photometric observations within the transit window are shown for each system in the lower
panels of Figures \ref{fig:HD9446_phase} and \ref{fig:HD43691b_phase}. The solid black line shows the transit baseline and the red line shows the binned average flux measurements for each planet. A dashed line shows the predicted transit signal expected for a central transit of the known planet. The vertical dotted lines give the $\pm 1 \sigma$ uncertainty in the timing of the transit window..

A comparison of the in-transit and out-of-transit flux for 
each system reveals there is no significant difference in the flux received during these times. When compared to the expected transit depths calculated in Section \ref{TP} for each system, we can rule out a central transit for planet HD~9446~b to a certainty of 3.96 sigma. Due to gaps in the photometry data we can only rule out a transit of HD~9446~c to a certainty of 3.88 sigma for a planet with an impact parameter of up to b~=~0.5778. 
Accounting for the data gap in the photometry of HD~43691 we can only rule out a transit to a certainty of 0.74 sigma for a transiting planet with an impact parameter of up to b~=~0.898.
For HD~179079~b a transit cannot be ruled out due to the combination of a relatively small planet with a large star and thus low signal to noise.
Note that for grazing transits, limb darkening will cause a reduction in the transit depth and so these types of transits have not been ruled out in our analysis \citep{Mandel2002}.

\clearpage

\subsection{Photometric Variability of Host Stars}
\label{PVHS} 

To check the photometric variability of all three host stars, we calculated and examined their Lomb--Scargle (L--S) periodograms \citep{Lomb76,Scargle82}. We calculated the L--S periodograms for each dataset using {\scriptsize{astropy}} \citep{Astropy13,Astropy18} for each observing season separately and also for the entire light curves, in the frequency range 0.01--1~cycles\,day$^{-1}$ with 10\,000 equally-spaced frequency steps. Frequencies above 0.98 were excluded from the analysis due to the strong periodicity peaks caused by the diurnal observing cycle.

In each periodogram, we identified the highest periodicity peak in the un-excluded frequency range as the candidate for periodic astrophysical variability. Next, we estimated the period uncertainty from the width of the periodicity peak, which we determined by fitting the peak with a Gaussian function in the frequency range equal to the inverse of the observing baseline centered at the peak’s central frequency. We also scaled the period uncertainty with the peak significance by dividing it with the square root of the normalized power of the peak \citep{VanderPlas18}. Then, we phase folded each light curve on the identified variability period, and determined the variability amplitude and its uncertainty by fitting the phase curve with a sine function. We also calculated the root mean square (RMS) for each light curve before and after the sinusoidal variability fit, and false alarm probability (FAP) for each periodogram and each highest peak using the Baluev approximation \citep{Baluev08}.

With the injection tests and by trial and error, we selected the following criteria for a statistically significant variability detection: (1) FAP of the periodicity peak has to be lower than 1\%, (2) the full-amplitude of the variability has to be greater than the RMS before variability fitting, and (3) the RMS has to drop by at least 10\% when removing the best-fit variability signal.

\begin{figure*}
\centering
\includegraphics[width=\textwidth]{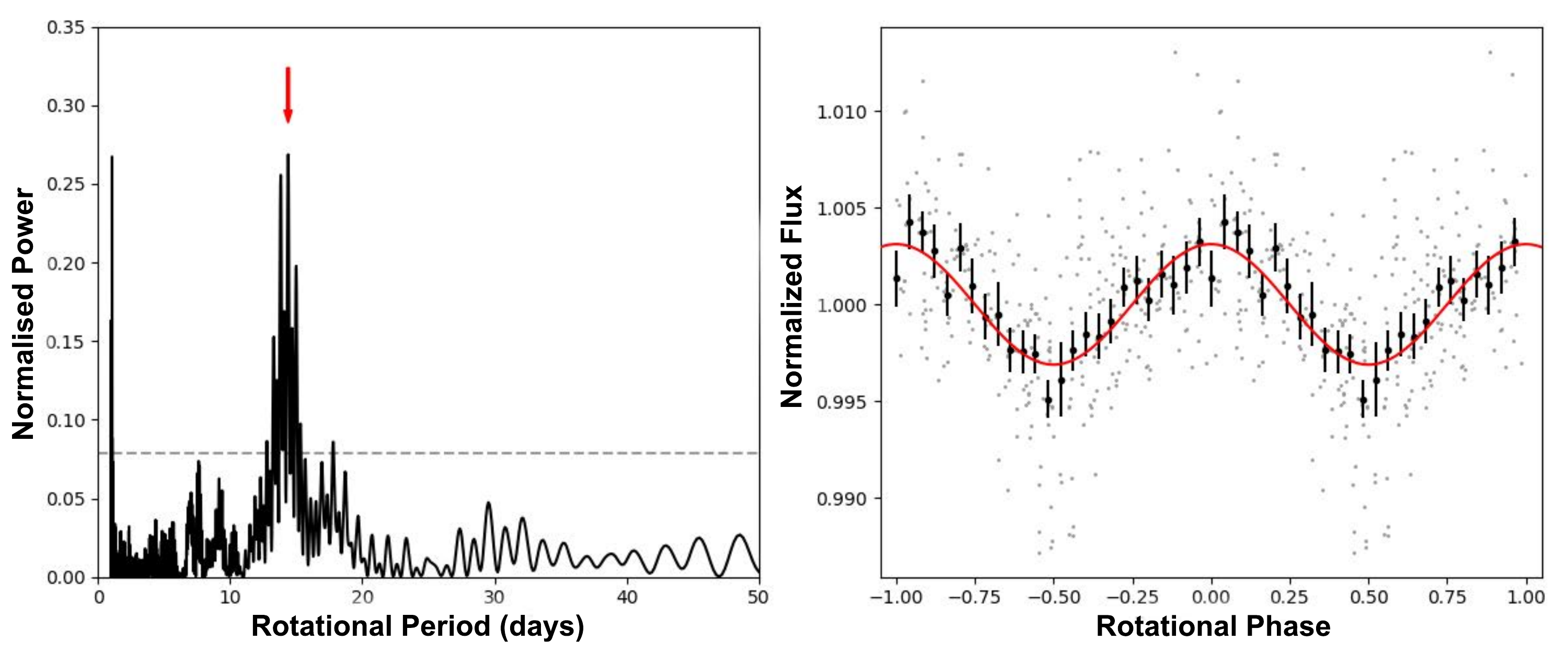}
\caption{Detection of the periodic variability of HD~9446 using the APT light curve. Left panel: L--S periodogram. The highest periodicity peak is marked with an arrow. The peak near 1-day is caused by the diurnal cycle. The horizontal dashed line represents the FAP of 1\%. Right panel: Rotational phase curve. Unbinned data points are shown with small grey symbols, which have been used in the variability analysis. Larger, black symbols with error bars represent the binned phase curve with 25 bins. We only used the unbinned data in our variability analysis, and the binned data are shown here only for the visualization purpose. The red line is the best sine fit to the unbinned rotational phase curve.}
\label{fig:HD9446_variability}
\end{figure*}

Following these criteria, we find that HD~9446 is variable with a period of $14.39\pm0.19$ days from our analysis of the whole APT light curve. We also detect the 14-day variability in every individual APT observing season (see Table~\ref{tab:HD9446_variability}). APT periodogram and variability phase curve of HD~9446 are shown in Fig.~\ref{fig:HD9446_variability}. Given the small amplitude of about half a percent, the sinusoidal behavior of the variability, and the G5V spectral type of the host star, we conclude that the 14-day periodicity detected in the APT data is likely caused by starspots. This agrees with the expected stellar rotational period of $12.5^{+4.1}_{-2.5}\sin i_*$ days, which was calculated from the stellar projected rotational velocity and stellar radius, and $\sim$10-day period, which is based on the measurement of the chromospheric activity index \citep{Hebrard10}.

\begin{table*}
\centering
\begin{minipage}{13.5cm}
\caption{Results of the periodicity analysis for HD~9446 using the APT data.}
\label{tab:HD9446_variability}
\begin{tabular}{ccccccc}
\hline
Season&Period [d]&Full Amplitude&L--S Power&FAP&RMS before&RMS after\\
\hline
2015-16&$14.7\pm2.1$&$0.0051\pm0.0010$&0.3026&0.0003&0.0032&0.0027\\
2016-17&$13.8\pm2.1$&$0.0037\pm0.0008$&0.2186&0.0095&0.0025&0.0022\\
2017-18&$14.1\pm1.4$&$0.0096\pm0.0013$&0.3846&$<$0.0001&0.0055&0.0044\\
all&$14.39\pm±0.19$&$0.0063\pm±0.0007$&0.2697&$<$0.0001&0.0043&0.0037\\
\hline
\end{tabular}
\end{minipage}
\end{table*}

We did not detect the 14-day periodicity in the KELT data, which may be due to a number of reasons. First, the redder and broader KELT passband filter compared to the APT is dampening more variability amplitude, which typically peaks in the blue part of the spectrum \citep{Oelkers18}. Second, the RMS of most KELT observing seasons was higher than the APTs. Third, since the KELT and APT datasets do not temporally overlap (see Table~\ref{tab:Photometric}), the starspots may have been absent or possibly distributed differently on the stellar surface during the KELT observations. However, the KELT periodograms of all observing seasons tentatively but consistently indicate a potential presence of around-7-day periodicity, which is compatible with half of the rotational period determined with APT, and might be interpreted as a bimodal distribution of starspots on the stellar surface. However, because the variability amplitudes indicated by the KELT are significantly lower than those detected in the APT data, and do not exceed our significance criteria, they should be regarded as tentative. Instead, we set a conservative upper limit for the variability amplitude of 0.005, equal to the RMS of the entire KELT light curve.

For HD~43691 and HD~179079, we did not detect any significant periodic variability in any of the APT or KELT whole or seasonal light curves. For HD~43691 we set the APT and KELT variability amplitude upper limits to 0.002 and 0.008, and for HD~179079 to 0.002 and 0.006, respectively.

Stellar chromospheric activity is tightly linked with the stellar magnetic field and is correlated with the magnetic cycle \citep{Hall08} and photometric rotational modulation \citep{Cau14}. We do not detect any periodicities in the Ca\,{\sc ii} H and K chromospheric activity indices within the available observational timespans of HIRES
observations (see Table~\ref{tab:Spectr})
for any of the three host stars. The corresponding $\log R^{'}_{\rm {HK}}$ measurements and their 1-$\sigma$ standard deviations for HD~9446, HD~43691, and HD~179079, are $-4.71\pm0.04$, $-4.85\pm0.02$, and $-5.00\pm0.01$, respectively.

\section{Conclusions}
\label{Conclu}

Using new RV data obtained from the Keck HIRES instrument we refined the orbital parameters of the known planets and looked for linear trends in the HD~9446, HD~179079 and HD~43691 systems. While our fits largely agreed with the published values for these systems, an improved solution for planet HD~9446~c was found, giving the outer planet an orbital period of 189.6$\pm 0.13$ days. We also found that for HD~9446 the best fit included a $\sim4.8\sigma$ trend. As the results from our speckle imaging analysis in Section \ref{SP} indicated no evidence of a stellar companion, this likely indicates an additional planet in the HD~9446 system and so further RV monitoring of this system should be prioritized. As the discovery paper by \citet{Hebrard10} did not report a linear trend for HD~9446, it is possible the original fit included some of the linear trend in the orbital solution for planet c.  The best fit for HD~43691 included a $\sim1.75\sigma$ trend. While this could also indicate the presence of another unknown planet in the system, this is a low sigma result and so warrants only a small mention here.

From the refined fits for each system we produced a transit ephemeris onto which we folded precise photometry acquired with the APTs and KELT.
For planet HD~9446~b a transit can be ruled out to a certainty of 3.96$\sigma$. For HD~9446~c we can rule out a transit to a certainty of 3.88$\sigma$ for a planet with an impact parameter of up to b~=~0.5778, and for planet HD~43691 a transit can be ruled out to a certainty of 0.74$\sigma$ for a planet with an impact parameter of up to b~=~0.898.
Each of these stars falls into the gaps of the primary planned mission for $TESS$ and so are unlikely to be observed by any space-based instruments. Thus these ground based observations are essential in determining the possible transit of the planets in these systems. Given the difficulties in ruling out a transit for HD~179079~b through ground based photometry, the total effective success probability
of the experiment is significantly lower than the transit probability described in Section \ref{TP}. Excluding HD~179079~b, the maximum success probability of the remaining planets is 6\% assuming HD~9446~b and HD~9446~c are co-planar, 5.3\% if not.

Our analysis of the T10 APT light curve revealed a $14.39\pm0.19$ day variability period of host star HD~9446.
We identify this as the rotational period of the host star.
Host stars HD~43691 and HD~179079 showed no evidence of significant periodic variability.

The TERMS project aids in the detection of planetary transits by refining the orbital parameters of known RV planets and calculating a revised transit ephemeris for the planet, and now combined with TESS photometry the expected yield of this collaboration is expected to be significant \citep{Dalba19}. Already there have been detections by TESS of known planets that utilised the ongoing RV observations made by the TERMS team \citep{Mocnik2020, Pepper2020}. Precision photometry from the {\it TESS} mission will be combined with the improved orbital ephemerides from TERMS observations to search for additional transiting planets in observed systems, such as the case of Pi Mensae c \citep{Huang18}. 
It is expected that many more systems observed by TERMS will have as-yet undetected additional planets that will be revealed through transits and trends seen in additional RV observations, such as that found in \citet{Wang2012}. TERMS observations also contribute to the stellar characterization of exoplanet host stars \citep{Dragomir2012}.

The high quality transit ephemerides
are essential for follow-up observations that will occur with future space missions. For example, transmission spectroscopic observations with the James Webb Space Telescope and direct imaging with missions such as WFIRST require that both the orbit of the planets within each system and the properties of the host star be determined to a high degree of certainty \citep{Beichman14,kane2018}. The refined orbital properties and transit ephemerides presented in this work will be used for observing strategy development of these systems for many years, and the ruling out of transits will allow for prioritization of other targets in future transit missions, saving valuable telescope time.
The targets studied by the TERMS project provide critical refinement of the orbital parameters of known planets and enable further observations to detect the existence of additional planets in orbit about each system observed.

\section*{Acknowledgements} 
Thanks to the anonymous referee, whose comments greatly improved the quality of the paper.
GWH acknowledges long-term support from NASA, NSF, Tennessee State University, 
and the State of Tennessee through its Centers of Excellence program. The research shown here acknowledges use of the Hypatia Catalog Database, an online compilation of stellar abundance data as described in \citet{Hinkel14}, which was supported by NASA's Nexus for Exoplanet System Science (NExSS) research coordination network and the Vanderbilt Initiative in Data-Intensive Astrophysics (VIDA). This research has made use of the NASA Exoplanet Archive, which is operated by the California Institute of Technology, under contract with the National Aeronautics and Space Administration under the Exoplanet Exploration Program. This research has made use of NASA’s Astrophysics Data System. This research has made use of the VizieR catalogue access tool, CDS, Strasbourg, France (DOI: 10.26093/cds/vizier). The original description of the VizieR service was published in A\&AS 143, 23

\newpage
\clearpage
\newpage
\clearpage

\appendix
\section{Radial Velocity Data}
\label{APP}

\startlongtable
\begin{deluxetable}{lrrc}
\tablecaption{ Radial Velocities HD~43691\tablenotemark{*}}
\tablehead{
  \colhead{Time} & 
  \colhead{RV} & 
  \colhead{RV Unc.} & 
  \colhead{Inst.} \\
  \colhead{(BJD\_TDB)} & 
  \colhead{(m s$^{-1}$)} & 
  \colhead{(m s$^{-1}$)} & 
  \colhead{}
}
\startdata
2453015.00100 & 158.82 & 2.15 & HIRES$_{\rm PRE}$$_{\rm 2004}$ \\
  2453015.99600 & 154.69 & 2.06 & HIRES$_{\rm PRE}$$_{\rm 2004}$ \\
  2453017.00100 & 155.06 & 2.05 & HIRES$_{\rm PRE}$$_{\rm 2004}$ \\
  2453071.81700 & -95.94 & 2.42 & HIRES$_{\rm PRE}$$_{\rm 2004}$ \\
  2453483.74700 & -90.71 & 1.64 & HIRES \\
  2453982.12600 & 142.31 & 1.35 & HIRES \\
  2455172.97600 & 17.66 & 1.52 & HIRES \\
  2455435.12900 & -32.43 & 1.20 & HIRES \\
  2455633.76700 & -24.75 & 1.54 & HIRES \\
  2455663.72500 & -102.27 & 1.56 & HIRES \\
  2455699.73300 & -107.72 & 1.62 & HIRES \\
  2455810.13200 & -102.60 & 1.34 & HIRES \\
  2455967.87200 & -9.54 & 1.52 & HIRES \\
  2455999.76400 & -91.19 & 1.65 & HIRES \\
  2456165.12600 & 140.90 & 1.43 & HIRES \\
  2456585.15400 & -78.78 & 1.24 & HIRES \\
  2456588.06400 & -100.76 & 1.18 & HIRES \\
  2456639.91200 & 136.12 & 1.61 & HIRES \\
  2456907.15100 & 107.44 & 1.26 & HIRES \\
  2457299.09500 & -16.06 & 1.51 & HIRES \\
  2457356.16900 & 15.93 & 1.49 & HIRES \\
  2457652.14700 & 12.98 & 1.48 & HIRES \\
  2457765.93800 & -31.24 & 1.36 & HIRES \\
  2458367.10100 & -110.55 & 1.30 & HIRES \\
  2458479.08400 & -99.33 & 1.48 & HIRES \\
  2458714.13600 & 151.17 & 1.45 & HIRES \\
  2458716.12900 & 145.80 & 1.34 & HIRES \\
  2458720.13800 & 104.91 & 1.37 & HIRES \\
  2458723.13800 & 53.19 & 1.38 & HIRES \\
  2458724.13400 & 38.51 & 1.25 & HIRES \\
  2458733.12100 & -92.44 & 1.27 & HIRES \\
\enddata
\tablenotetext{*}{In addition, this study used the RV data from SOPHIE and ELODIE published in \citet{Dasilva07}}
\end{deluxetable}

\newpage
\startlongtable
\begin{deluxetable}{lrrc}
\tablecaption{ Radial Velocities HD~9446\tablenotemark{*} }
\tablehead{
  \colhead{Time} & 
  \colhead{RV} & 
  \colhead{RV Unc.} & 
  \colhead{Inst.} \\
  \colhead{(BJD\_TDB)} & 
  \colhead{(m s$^{-1}$)} & 
  \colhead{(m s$^{-1}$)} & 
  \colhead{}
}
\startdata
  2456588.95400 & -19.84 & 1.44 & HIRES \\
  2456708.72400 & 33.14 & 1.59 & HIRES \\
  2456844.12600 & -42.03 & 1.24 & HIRES \\
  2456849.12800 & -2.48 & 1.57 & HIRES \\
  2456861.05200 & -13.62 & 1.66 & HIRES \\
  2456867.09011 & -50.61 & 0.90 & HIRES \\
  2456882.11200 & 57.61 & 1.55 & HIRES \\
  2456910.11000 & 97.61 & 1.72 & HIRES \\
  2457001.76600 & -43.22 & 1.47 & HIRES \\
  2457058.71700 & 24.48 & 1.59 & HIRES \\
  2457237.13500 & 6.41 & 1.23 & HIRES \\
  2457240.12500 & 37.39 & 1.30 & HIRES \\
  2457265.11900 & 67.90 & 1.47 & HIRES \\
  2457291.00200 & 23.56 & 1.53 & HIRES \\
  2457582.13400 & -77.37 & 1.55 & HIRES \\
  2457587.11200 & -120.59 & 1.45 & HIRES \\
  2457595.03900 & -21.99 & 1.42 & HIRES \\
  2457791.73900 & -59.49 & 1.91 & HIRES \\
  2457802.70400 & -51.55 & 1.67 & HIRES \\
  2457925.12800 & -31.16 & 1.77 & HIRES \\
  2457995.00500 & 27.36 & 1.64 & HIRES \\
  2457999.97500 & 19.30 & 1.71 & HIRES \\
  2458000.97500 & -6.74 & 1.54 & HIRES \\
  2458003.00600 & -19.13 & 1.71 & HIRES \\
  2458023.94400 & 74.72 & 1.69 & HIRES \\
  2458028.90700 & 52.95 & 1.79 & HIRES \\
  2458029.87500 & 39.92 & 1.83 & HIRES \\
  2458065.81000 & 11.05 & 1.88 & HIRES \\
  2458112.69300 & 15.22 & 1.64 & HIRES \\
  2458337.12000 & -80.41 & 1.72 & HIRES \\
  2458341.10700 & -93.33 & 1.58 & HIRES \\
  2458364.03100 & -42.91 & 1.61 & HIRES \\
  2458385.99900 & 40.13 & 1.56 & HIRES \\
  2458394.03500 & 24.10 & 1.75 & HIRES \\
  2458414.89600 & 74.24 & 1.63 & HIRES \\
  2458426.97000 & 36.00 & 1.81 & HIRES \\
  2458462.74500 & 3.38 & 1.64 & HIRES \\
  2458476.78700 & 41.47 & 1.78 & HIRES \\
  2458714.03700 & -11.36 & 1.24 & HIRES \\
  2458715.05300 & -5.62 & 1.43 & HIRES \\
  2458716.04400 & -3.44 & 1.22 & HIRES \\
  2458720.03800 & -51.58 & 1.28 & HIRES \\
  2458723.03000 & -74.25 & 1.27 & HIRES \\
  2458724.04400 & -71.00 & 1.29 & HIRES \\
  2458737.88700 & -16.74 & 1.91 & HIRES \\
  2458739.02800 & 0.23 & 1.69 & HIRES \\
  2458739.93500 & 8.59 & 1.69 & HIRES \\
  2458742.98100 & 23.66 & 1.69 & HIRES \\
  2458743.97100 & 31.39 & 1.64 & HIRES \\
  2458746.97100 & 7.93 & 1.64 & HIRES \\
  2458774.89000 & 40.19 & 1.77 & HIRES \\
\enddata
\tablenotetext{*}{In addition, this study used the RV data from SOPHIE published in \citet{Hebrard10}.}
\end{deluxetable}

\startlongtable
\begin{deluxetable}{lrrc}
\tablecaption{ Radial Velocities HD~179079 }
\tablehead{
  \colhead{Time} & 
  \colhead{RV} & 
  \colhead{RV Unc.} & 
  \colhead{Inst.} \\
  \colhead{(BJD\_TDB)} & 
  \colhead{(m s$^{-1}$)} & 
  \colhead{(m s$^{-1}$)} & 
  \colhead{}
}
\startdata
2453197.997 & 5.02 & 2.21 & HIRES \\
2453198.963 & 8.11 & 2.96 & HIRES \\
2453199.909 & 1.63 & 2.45 & HIRES \\
2453208.025 & -14.32 & 2.1 & HIRES \\
2453603.86 & 8.87 & 1.57 & HIRES \\
2453961.876 & -10.82 & 1.39 & HIRES \\
2453963.867 & -3.23 & 1.41 & HIRES \\
2453981.76 & 2.18 & 1.38 & HIRES \\
2453982.806 & 3.3 & 1.43 & HIRES \\
2453983.77 & 0.82 & 1.34 & HIRES \\
2453984.845 & -2.38 & 1.27 & HIRES \\
2454249.037 & -13.09 & 1.28 & HIRES \\
2454250.08 & -13.21 & 1.33 & HIRES \\
2454251.057 & -6.49 & 1.08 & HIRES \\
2454251.936 & -3.46 & 1 & HIRES \\
2454256.09 & 5.71 & 1.12 & HIRES \\
2454279.039 & -5.75 & 1.07 & HIRES \\
2454280.047 & -9.91 & 1.16 & HIRES \\
2454286.038 & 0.26 & 1.36 & HIRES \\
2454304.972 & -2.15 & 1.19 & HIRES \\
2454305.972 & -8.05 & 1.17 & HIRES \\
2454306.972 & 0.12 & 1.16 & HIRES \\
2454308.001 & -1.76 & 1.24 & HIRES \\
2454308.969 & 0.43 & 1.18 & HIRES \\
2454309.965 & -1.7 & 1.2 & HIRES \\
2454310.957 & 5.17 & 1.15 & HIRES \\
2454311.955 & 5.03 & 1.21 & HIRES \\
2454312.95 & 3.42 & 1.28 & HIRES \\
2454313.948 & 5.51 & 1.16 & HIRES \\
2454314.957 & 12.62 & 1.07 & HIRES \\
2454318.863 & -7.87 & 1.15 & HIRES \\
2454335.965 & -3.47 & 1.23 & HIRES \\
2454336.989 & 4.31 & 1.32 & HIRES \\
2454339.853 & 5.58 & 1.1 & HIRES \\
2454343.888 & 3.48 & 1.18 & HIRES \\
2454344.944 & -5.03 & 1.21 & HIRES \\
2454345.759 & -7.69 & 1.18 & HIRES \\
2454396.73 & -2.48 & 1.24 & HIRES \\
2454397.757 & -3.77 & 1.13 & HIRES \\
2454398.742 & -0.45 & 1.12 & HIRES \\
2454399.725 & -1.18 & 1.35 & HIRES \\
2454427.745 & -0.23 & 1.19 & HIRES \\
2454428.704 & 1.42 & 1.19 & HIRES \\
2454429.686 & 2.22 & 1.18 & HIRES \\
2454430.683 & -1.02 & 1.23 & HIRES \\
2454548.152 & -1.25 & 1.23 & HIRES \\
2454549.145 & -5.61 & 1.26 & HIRES \\
2454602.977 & 3.27 & 1.17 & HIRES \\
2454603.997 & 12.27 & 1.43 & HIRES \\
2454634.049 & 0.33 & 1.36 & HIRES \\
2454634.978 & 1.22 & 1.25 & HIRES \\
2454636.021 & -2.47 & 1.21 & HIRES \\
2454637.067 & -6.63 & 1.26 & HIRES \\
2454638.014 & -7.45 & 1.2 & HIRES \\
2454639.045 & -9.97 & 1.23 & HIRES \\
2454640.127 & -11.82 & 1.35 & HIRES \\
2454641.006 & -10.92 & 1.19 & HIRES \\
2454642.104 & -7.72 & 1.28 & HIRES \\
2454644.1 & -0.49 & 1.23 & HIRES \\
2454674.838 & 5.57 & 1.16 & HIRES \\
2454688.849 & 5.49 & 1.25 & HIRES \\
2454690.022 & 3.18 & 1.34 & HIRES \\
2454717.772 & 0.94 & 1.16 & HIRES \\
2454718.791 & 4.69 & 1.13 & HIRES \\
2454719.803 & 3.68 & 1.15 & HIRES \\
2454720.842 & 5.43 & 1.18 & HIRES \\
2454721.829 & -4.16 & 1.16 & HIRES \\
2454722.772 & -0.6 & 1.19 & HIRES \\
2454723.765 & -2.57 & 1.23 & HIRES \\
2454724.779 & -3.78 & 1.22 & HIRES \\
2454725.77 & -11.84 & 1.29 & HIRES \\
2454726.766 & -9.1 & 1.1 & HIRES \\
2454727.843 & -12.92 & 1.21 & HIRES \\
2454777.763 & 3.5 & 1.33 & HIRES \\
2455173.69 & -8.87 & 1.33 & HIRES \\
2455319.076 & -7.09 & 1.29 & HIRES \\
2455395.94 & 0.52 & 1.26 & HIRES \\
2455490.74 & 5.34 & 1.25 & HIRES \\
2455636.122 & -7.95 & 1.13 & HIRES \\
2455671.065 & 2.65 & 1.17 & HIRES \\
2455796.777 & -11.46 & 1.3 & HIRES \\
2456195.751 & 2.78 & 1.25 & HIRES \\
2456637.686 & -10.66 & 1.6 & HIRES \\
2456908.727 & 3.82 & 1.11 & HIRES \\
2457215.937 & 3.98 & 1.38 & HIRES \\
2457285.742 & -6.67 & 1.26 & HIRES \\
2457831.152 & -1.65 & 1.21 & HIRES \\
2458346.797 & -5.47 & 1.45 & HIRES \\
2458713.775 & 1.47 & 1.26 & HIRES \\
2458714.745 & 5.62 & 1.24 & HIRES \\
2458715.741 & -1.67 & 1.26 & HIRES \\
2458722.768 & 3.01 & 1.16 & HIRES \\
2458723.741 & 0.43 & 1.22 & HIRES \\
\enddata
\end{deluxetable}

\section{RadVel Model Comparison Tables}
\label{MCT}


\begin{deluxetable*}{llrrrrrrr}
\tablecaption{Model Comparison HD~9446\label{tab:BIC_9446}}
\tablehead{\colhead{AICc Qualitative Comparison} & \colhead{Free Parameters} & \colhead{$N_{\rm free}$} & \colhead{$N_{\rm data}$} & \colhead{RMS} & \colhead{$\ln{\mathcal{L}}$} & \colhead{BIC} & \colhead{AICc} & \colhead{$\Delta$AICc}}
\startdata
  AICc Favored Model & $e_{b}$, $K_{b}$, $e_{c}$, $K_{c}$, $\dot{\gamma}$, {$\sigma$}, {$\gamma$} & 13 & 129 & 14.16 & -522.62 & 1098.67 & 1064.66 & 0.00 \\
  \hline \\
  Nearly Indistinguishable & $K_{b}$, $e_{c}$, $K_{c}$, $\dot{\gamma}$, {$\sigma$}, {$\gamma$} & 11 & 129 & 14.50 & -526.08 & 1095.24 & 1066.04 & 1.38 \\
  \hline \\
  Ruled Out & $K_{b}$, $e_{c}$, $K_{c}$, {$\sigma$}, {$\gamma$} & 10 & 129 & 15.54 & -536.56 & 1110.87 & 1084.14 & 19.48 \\
   & $e_{b}$, $K_{b}$, $e_{c}$, $K_{c}$, {$\sigma$}, {$\gamma$} & 12 & 129 & 15.29 & -534.51 & 1117.00 & 1085.37 & 20.71 \\
   & $e_{b}$, $K_{b}$, $K_{c}$, $\dot{\gamma}$, {$\sigma$}, {$\gamma$} & 11 & 129 & 15.79 & -536.49 & 1115.84 & 1086.64 & 21.98 \\
   & $K_{b}$, $K_{c}$, $\dot{\gamma}$, {$\sigma$}, {$\gamma$} & 9 & 129 & 16.26 & -540.43 & 1113.33 & 1089.11 & 24.45 \\
   & $e_{b}$, $K_{b}$, $K_{c}$, {$\sigma$}, {$\gamma$} & 10 & 129 & 16.44 & -543.58 & 1124.53 & 1097.79 & 33.13 \\
    & $K_{b}$, $K_{c}$, {$\sigma$}, {$\gamma$} & 8 & 129 & 16.88 & -546.73 & 1120.75 & 1099.07 & 34.41 \\
   & $K_{b}$, $\dot{\gamma}$, {$\sigma$}, {$\gamma$} & 6 & 129 & 33.75 & -636.54 & 1292.58 & 1276.11 & 211.45 \\
   & $e_{b}$, $K_{b}$, $\dot{\gamma}$, {$\sigma$}, {$\gamma$} & 8 & 129 & 33.32 & -635.00 & 1298.24 & 1276.57 & 211.91 \\
   & $K_{b}$, {$\sigma$}, {$\gamma$} & 5 & 129 & 34.06 & -638.04 & 1290.49 & 1276.68 & 212.02 \\
   & $e_{b}$, $K_{b}$, {$\sigma$}, {$\gamma$} & 7 & 129 & 33.74 & -637.10 & 1297.53 & 1278.44 & 213.78 \\
   & $K_{c}$, $\dot{\gamma}$, {$\sigma$}, {$\gamma$} & 6 & 129 & 43.66 & -668.97 & 1357.00 & 1340.53 & 275.87 \\
   & $K_{c}$, {$\sigma$}, {$\gamma$} & 5 & 129 & 44.08 & -670.22 & 1354.68 & 1340.87 & 276.21 \\
   & $e_{c}$, $K_{c}$, {$\sigma$}, {$\gamma$} & 7 & 129 & 43.87 & -669.74 & 1363.49 & 1344.40 & 279.74 \\
   & $e_{c}$, $K_{c}$, $\dot{\gamma}$, {$\sigma$}, {$\gamma$} & 8 & 129 & 43.53 & -668.69 & 1366.22 & 1344.54 & 279.88 \\
   & {$\sigma$}, {$\gamma$} & 2 & 129 & 53.53 & -694.73 & 1389.28 & 1383.66 & 319.00 \\
   & $\dot{\gamma}$, {$\sigma$}, {$\gamma$} & 3 & 129 & 53.20 & -694.03 & 1392.73 & 1384.35 & 319.69 \\
   & $e_{b}$, $K_{b}$, $e_{c}$, $K_{c}$, $\dot{\gamma}$, {$\gamma$} & 11 & 129 & 14.59 & -1720.01 & 3483.12 & 3453.92 & 2389.26 \\
   & $K_{b}$, $e_{c}$, $K_{c}$, $\dot{\gamma}$, {$\gamma$} & 9 & 129 & 15.07 & -1806.56 & 3646.11 & 3621.89 & 2557.23 \\
   & $e_{b}$, $K_{b}$, $K_{c}$, $\dot{\gamma}$, {$\gamma$} & 9 & 129 & 16.26 & -2016.26 & 4064.95 & 4040.73 & 2976.07 \\
   & $K_{b}$, $K_{c}$, $\dot{\gamma}$, {$\gamma$} & 7 & 129 & 16.76 & -2101.19 & 4224.76 & 4205.66 & 3141.00 \\
   & $e_{b}$, $K_{b}$, $e_{c}$, $K_{c}$, {$\gamma$} & 10 & 129 & 16.11 & -2141.88 & 4318.99 & 4292.26 & 3227.60 \\
   & $K_{b}$, $e_{c}$, $K_{c}$, {$\gamma$} & 8 & 129 & 16.18 & -2179.70 & 4384.89 & 4363.21 & 3298.55 \\
   & $e_{b}$, $K_{b}$, $K_{c}$, {$\gamma$} & 8 & 129 & 17.38 & -2352.36 & 4729.83 & 4708.15 & 3643.49 \\
   & $K_{b}$, $K_{c}$, {$\gamma$} & 6 & 129 & 17.65 & -2412.14 & 4839.35 & 4822.88 & 3758.22 \\
   & $e_{b}$, $K_{b}$, $\dot{\gamma}$, {$\gamma$} & 6 & 129 & 34.00 & -11728.06 & 23472.15 & 23455.67 & 22391.01 \\
   & $K_{b}$, $\dot{\gamma}$, {$\gamma$} & 4 & 129 & 33.74 & -12298.96 & 24604.46 & 24593.34 & 23528.68 \\
   & $e_{b}$, $K_{b}$, {$\gamma$} & 5 & 129 & 35.59 & -12696.38 & 25397.32 & 25383.51 & 24318.85 \\
   & $K_{b}$, {$\gamma$} & 3 & 129 & 35.02 & -13079.73 & 26156.88 & 26148.49 & 25083.83 \\
   & $e_{c}$, $K_{c}$, $\dot{\gamma}$, {$\gamma$} & 6 & 129 & 44.71 & -17516.10 & 35051.55 & 35035.08 & 33970.42 \\
   & $K_{c}$, $\dot{\gamma}$, {$\gamma$} & 4 & 129 & 44.42 & -17565.82 & 35141.27 & 35130.16 & 34065.50 \\
   & $e_{c}$, $K_{c}$, {$\gamma$} & 5 & 129 & 45.43 & -18105.53 & 36225.48 & 36211.66 & 35147.00 \\
   & $K_{c}$, {$\gamma$} & 3 & 129 & 45.07 & -18160.39 & 36325.53 & 36317.14 & 35252.48 \\
   & $\dot{\gamma}$, {$\gamma$} & 1 & 129 & 53.39 & -27445.77 & 54886.50 & 54883.67 & 53819.01 \\
   & {$\gamma$} & 0 & 129 & 53.60 & -27570.93 & 55131.95 & 55131.95 & 54067.29 \\
\enddata
\end{deluxetable*}


\begin{deluxetable*}{llrrrrrrr}
\tablecaption{Model Comparison HD~43691 \label{tab:BIC_43691}}
\tablehead{\colhead{AICc Qualitative Comparison} & \colhead{Free Parameters} & \colhead{$N_{\rm free}$} & \colhead{$N_{\rm data}$} & \colhead{RMS} & \colhead{$\ln{\mathcal{L}}$} & \colhead{BIC} & \colhead{AICc} & \colhead{$\Delta$AICc}}
\startdata
  AICc Favored Model & $e_{b}$, $K_{b}$, $\dot{\gamma}$, {$\sigma$}, {$\gamma$} & 10 & 67 & 11.47 & -224.05 & 501.61 & 483.49 & 0.00 \\
  \hline \\
  Nearly Indistinguishable & $e_{b}$, $K_{b}$, {$\sigma$}, {$\gamma$} & 9 & 67 & 11.48 & -226.02 & 501.65 & 484.96 & 1.47 \\
  \hline \\
  Ruled Out & $K_{b}$, {$\sigma$}, {$\gamma$} & 7 & 67 & 13.41 & -245.37 & 531.91 & 518.38 & 34.89 \\
   & $K_{b}$, $\dot{\gamma}$, {$\sigma$}, {$\gamma$} & 8 & 67 & 13.37 & -244.83 & 535.08 & 519.92 & 36.43 \\
   & {$\sigma$}, {$\gamma$} & 4 & 67 & 89.30 & -373.15 & 774.61 & 766.44 & 282.95 \\
   & $\dot{\gamma}$, {$\sigma$}, {$\gamma$} & 5 & 67 & 88.39 & -372.45 & 777.42 & 767.38 & 283.89 \\
\enddata
\end{deluxetable*}

\begin{deluxetable*}{llrrrrrrr}
\tablecaption{Model Comparison HD~179079 \label{tab:BIC_179079}}
\tablehead{\colhead{AICc Qualitative Comparison} & \colhead{Free Parameters} & \colhead{$N_{\rm free}$} & \colhead{$N_{\rm data}$} & \colhead{RMS} & \colhead{$\ln{\mathcal{L}}$} & \colhead{BIC} & \colhead{AICc} & \colhead{$\Delta$AICc}}
\startdata
  AICc Favored Model & $K_{b}$, {$\sigma$}, {$\gamma$} & 5 & 93 & 4.19 & -268.46 & 549.03 & 537.06 & 0.00 \\
  \hline \\
  Nearly Indistinguishable & $K_{b}$, $\dot{\gamma}$, {$\sigma$}, {$\gamma$} & 6 & 93 & 4.16 & -267.75 & 552.14 & 537.93 & 0.87 \\
  \hline \\
  Ruled Out & {$\sigma$}, {$\gamma$} & 2 & 93 & 6.06 & -301.06 & 600.65 & 595.72 & 58.66 \\
   & $\dot{\gamma}$, {$\sigma$}, {$\gamma$} & 3 & 93 & 6.05 & -300.95 & 604.96 & 597.63 & 60.57 \\
\enddata
\end{deluxetable*}

\newpage
\clearpage


\end{document}